\RequirePackage{fix-cm}
\documentclass[natbib,smallextended]{svjour3}       % onecolumn (second format)
\smartqed  % flush right qed marks, e.g. at end of proof
\usepackage{graphicx}
\usepackage{amsmath}
\usepackage{subfigure}
\usepackage{listings} % 

 \journalname{Transport in Porous Media}

\RequirePackage[top=2.0cm,bottom=2.4cm,left=2.4cm,right=2.4cm,includehead,includefoot]{geometry}

\begin{document}

\title{A numerical study of two-phase flow models with dynamic capillary pressure and hysteresis
%\thanks{}
}
% Grants or other notes about the article that should go on the front
% page should be placed within the \thanks{} command in the title
% (and the %-sign in front of \thanks{} should be deleted)
%
% General acknowledgments should be placed at the end of the article.

%\subtitle{Do you have a subtitle?\\ If so, write it here}

%\titlerunning{Short form of title}        % if too long for running head

\author{Hong Zhang \and
        Paul Andries Zegeling %etc.
}

%\authorrunning{Short form of author list} % if too long for running head

\institute{Corresponding author: H. Zhang.
            \\ H. Zhang \at
              \email{h.zhang4@uu.nl}            \\
            Department of Mathematics, Faculty of Science, Utrecht University, Budapestlaan 6, 3584CD Utrecht, The Netherlands
%             \emph{Present address:} of F. Author  %  if needed
           \and
           P. A. Zegeling \at
            \email{p.a.zegeling@uu.nl}\\
            Department of Mathematics, Faculty of Science, Utrecht University, Budapestlaan 6, 3584CD Utrecht, The Netherlands
}

\date{Received: date / Accepted: date}
% The correct dates will be entered by the editor

\maketitle

\begin{abstract}
Saturation overshoot and pressure overshoot are studied by incorporating dynamic capillary pressure, capillary pressure hysteresis and hysteretic dynamic coefficient with a traditional fractional flow equation in one dimension space. Using the method of lines, the discretizations are constructed by applying the Castillo-Grone's mimetic operators in the space direction and a semi-implicit integrator in the time direction. Convergence tests and conservation properties of the schemes are presented. Computed profiles capture both the saturation overshoot and pressure overshoot phenomena. Comparisons between numerical results and experiments illustrate the effectiveness and different features of the models.
% \PACS{PACS code1 \and PACS code2 \and more}
% \subclass{MSC code1 \and MSC code2 \and more}
\end{abstract}

\section{Introduction}
Water infiltrating into initially dry sandy porous media has been shown to produce saturation overshoot and pressure overshoot in \cite{selker1992fingered,shiozawa2004unexpected,dicarlo2004experimental,dicarlo2007capillary}. \cite{eliassi2001continuum} and \cite{egorov2003stability} have demonstrated that the traditional Richards equation is unable to describe saturation overshoot. To describe the non-monotonic behaviour, various extensions to the Richards equation have been investigated. \cite{eliassi2001continuum} studied three additional forms referred to as hypodiffusive form, hyperbolic form and mixed form, saturation overshoot are obtained by using the hypodiffusive form in \cite{eliassi2003porous}. \cite{dicarlo2008nonmonotonic} studied a non-monotonic capillary pressure-saturation relationship and a second order hyperbolic term, but they mentioned these extensions need a regularization term to produce a unique solution. \cite{cueto2009phase} explained the formation of gravity fingers during water infiltration in soil by introducing a fourth-order term to Richards equation. The tip and tail saturations after their phase model have good agreement with the experiments in \cite{dicarlo2004experimental}. \cite{nieber2003non} obtained non-monotonic saturation profiles by supplementing the Richards equation with a non-equilibrium capillary pressure-saturation relationship, as well as including hysteretic effects. Later, \cite{chapwanya2010numerical} studied gravity-driven fingering instabilities based on the work of \cite{nieber2003non}, their results demonstrate that the non-equilibrium Richards equation is capable of reproducing realistic fingering flows for a wide range of physically relevant parameters. 

Besides extensions to the Richards equation, other approaches to characterize the saturation overshoot have also been proposed. Such as the generalized theory by introducing percolating and non-percolating fluid phases into the traditional mathematical model \cite{hilfer2000macroscopic,hilfer2012nonmonotone,doster2010numerical}, fractional flow approach \cite{dicarlo2012fractional} and moment analysis \cite{xiong2012moment}.

Among all the proposed theories and models, the dynamic (or non-equilibrium) capillary pressure relationship proposed by \cite{stauffer1978time}, \cite{hassanizadeh1993thermodynamic}, \cite{kalaydjian1992dynamic} has received much attention. Results on travelling wave solutions, global existence, phase plane analysis and uniqueness of weak solutions are given in \cite{cuesta2000infiltration,van2007new,van2013travelling,mikelic2010global,spayd2011buckley,cao2015uniqueness}. In order to provide accurate simulations, several numerical methods have been proposed in literature, including a finite difference method with minmod slope limiter in \cite{van2007new}, a cell-centered finite difference method and a locally conservative Eulerian-Lagrangian method in \cite{peszynska2008numerical}, Godunov-type staggered central schemes in \cite{wang2013central}, two semi-implicit schemes based on equivalent reformulations in \cite{fan2013equivalent}, an adaptive moving mesh method in \cite{zegeling2015adaptive} and the fast explicit operator splitting method in \cite{kao2015fast}. 

%Following this idea, \cite{nieber2003non} used different values of $\tau$ in the wetting stage and the drainage stage.
In the previous studies of saturation overshoot, the dynamic capillary pressure model with a constant dynamic coefficient usually brought oscillations behind the drainage front \cite{dicarlo2005modeling,sander2008dynamic,van2013non}. The hysteretic non-equilibrium model proposed in \cite{beliaev2001theoretical} postulates that the dynamic capillary effects are significant only outside the main hysteresis loop. Following this idea, \cite{nieber2003non} adopted a saturation and pressure dependent dynamic coefficient $\tau = \tau_s^0 P'_w(s) (p_0 - p)_+^\gamma$. In this treatment, the Mualem hysteresis model was restricted only to the two-stage wetting-drainage process: trajectories for the wetting stage was located within $H_w$ (domain above the main wetting curve), while trajectories for the drainage stage was limited to $H_0$ (main hysteresis loop region). In the wetting stage $\tau_s^0 = \tau_w$ and in the drainage stage $\tau_s^0 = 0$ (in the numerical simulation the value was a small constant $=10^{-3}$).  Recently, the hysteretic dynamic capillary pressure effect has been reported in \cite{sakaki2010direct}. \cite{mirzaei2013experimental} shows that the dynamic effect in the relationship between capillary pressure and saturation is hysteretic in nature. In this contribution, we consider the hysteresis effects in capillary pressure and study the saturation overshoot and pressure overshoot by adding dynamic capillary pressure, capillary pressure hysteresis and hysteretic dynamic coefficient to a traditional fractional flow equation.

Two-phase flow models in porous media usually consist of coupled, nonlinear partial differential equations. Many reliable discretizations have been proposed for the Richards equation, for instance, the Galerkin finite elements in \cite{arbogast1993numerical}, the multipoint flux approximation in \cite{klausen2008convergence} and so on. The space and time discretizations usually lead to a large system of nonlinear equations, which makes the numerical simulation of two-phase flow a challenging task. Some popular methods used for the Richards equation are the Newton method \cite{radu2006convergence}, Picard method \cite{zarba1990general} or L-scheme \cite{radu2015robust}, for an extensive review we refer to \cite{list2016study}. The appearance of the dynamic capillary pressure term in the two-phase flow model adds additional difficulty to the numerical treatment. In \cite{sander2008dynamic}, a reformulation of the non-equilibrium two-phase flow equation, which consists of an elliptic equation and an ODE, is shown to be effective for numerical simulations, \cite{fan2013equivalent} also presented a reliable and efficient semi-implicit scheme for a similar form. In this paper, we present our schemes based on this reformulation. Because of the conservation property and easy implementation of the Castillo Grone's mimetic (CGM) operators, we apply the mimetic finite difference method to the elliptic equation in the space direction, then integrate the system in the time direction with the implicit trapezoidal rule.

The rest of the paper is organized as follows. In section 2, we first derive the traditional equation for one-dimensional two-phase flow, and then we present the extended models by incorporating the dynamic capillary pressure term and hysteresis effects. Section 3 is devoted to presenting the CGM operators. In section 4, we apply the CGM operators in the space direction and an implicit trapezoidal integrator in the time direction to discretize the system. In section 5, numerical experiments are carried out to show the effectiveness and reliability of the extended models. Section 6 summarizes the conclusions.

\section{Mathematical  Models}\label{sec:models}
Two-phase flow in porous media can be characterized by the saturation and pressure in each phase. The saturation in each phase is defined as the fraction of the pore volume occupied by the phase and is denoted as $S_\alpha$ where $\alpha = w, n$ is an index for wetting and non-wetting phases. For the derivation of the fractional flow equation we refer to \cite{hilfer2014saturation}. Let the gravity act in the positive $x$-direction, for each phase, the mass conservation law is represented by the equation
\begin{align}\label{eqn:masscons}
	\frac{\partial (\phi \rho_\alpha S_\alpha)}{\partial t} + \frac{\partial}{\partial x} (\rho_\alpha v_\alpha) = \rho_\alpha F_\alpha, \quad \alpha = n, w,
\end{align}
where $\phi$ is the porosity of the porous medium, $\rho_\alpha$, $v_\alpha$ and $F_\alpha$ are the density, volumetric velocity and source of each phase.

In Darcy scale, the balance of momentum of each phase is given by the Darcy's law
\begin{align}
	v_\alpha & = - \frac{k_{r \alpha} K }{\mu_\alpha} \frac{\partial}{\partial x} (p_\alpha - \rho_\alpha g x) \nonumber \\
	& = - \lambda_\alpha (\frac{\partial p_\alpha}{\partial x} - \rho_\alpha g),\quad \alpha = n, w, \label{eqn:velo}
\end{align}
where $K$ is the intrinsic permeability of the porous medium, $g$ is the gravitational acceleration constant, $k_{r \alpha}$, $\mu_\alpha$, $\lambda_\alpha = \frac{k_{r \alpha} K}{\mu_\alpha}$ and $p_\alpha$ are the relative permeability function, viscosity, mobility and pressure of phase $\alpha$, respectively.

For the two-phase system the following constitutive relation holds:
\begin{align}\label{eqn:totasatu}
  S_w + S_n = 1.
\end{align}
Then $k_{rn}$ and $ \lambda_{n}$ can be written as functions of $S_w$. The total velocity is given by
\begin{align}\label{eqn:totalvelo}
  v_T = v_n + v_w.
\end{align}
Using Eqs. (\ref{eqn:velo}), (\ref{eqn:totasatu}) and (\ref{eqn:totalvelo}) the velocity of the wetting phase can be expressed in terms of the phase mobilities, total velocity and phases pressure difference as
\begin{align}\label{eqn:vwfrac}
	v_w = v_T \frac{\lambda_w}{\lambda_T}[ 1 + \frac{\lambda_n}{v_T} (\frac{\partial}{\partial x} (p_n - p_w) + (\rho_w - \rho_n) g)].
\end{align}
Assuming $\phi$ and temperature are constant, the phases are incompressible, by substituting Eq. (\ref{eqn:vwfrac}) into Eq. (\ref{eqn:masscons}) for the wetting phase, we obtain a nonlinear equation for the wetting phase
\begin{equation}\label{eqn:equi0}
	\phi \frac{\partial {S}_w}{\partial t} +  \frac{\partial}{\partial x} \big[v_T \frac{\lambda_w}{\lambda_T}[ 1 + \frac{\lambda_n}{v_T} (\frac{\partial}{\partial x} (p_n - p_w) + (\rho_w - \rho_n)g )]\big] =  F_w.
\end{equation}

In \cite{dicarlo2004experimental} the experiments were conducted in thin tubes, and water was injected into the tubes with different flux rates. Since the medium is homogeneous and the initial saturation is constant, the experiments are viewed as one-dimensional. Reminding that our aim is to simulate these experiments, we set the source term $F_w = 0$ and consider a flux boundary condition at $x_L$ and a Dirichlet boundary condition at $x_R$. Let $f(S_w) = \frac{\lambda_w}{\lambda_w + \lambda_n} = \frac{\lambda_w}{\lambda_T}$, then we have
\begin{align}
  \label{eqn:equi}
	\phi \frac{\partial{S}_w}{\partial t} + \frac{\partial}{\partial x} \big[q f(S_w)  + \lambda_n(S_w) f(S_w) \big(\frac{\partial}{\partial x} (p_n - p_w) + (\rho_w - \rho_n) g\big)\big] = 0,
\end{align}
with boundary conditions
\begin{equation}
  \label{eqn:bc}
  \left\{ \begin{array}{ll}
  v_w = q f(S_w) + \lambda_n(S_w)f(S_w)[ \frac{\partial }{\partial x} (p_n - p_w)+ (\rho_w - \rho_n) g] = q, & \mathrm{at~}x_L,\\
  S_w = S_{w}^{R}, & \mathrm{at~}x_R,
  \end{array}\right.
\end{equation}
where $q$ is the flux value used in \cite{dicarlo2004experimental,dicarlo2007capillary}, $S_{w}^{R}$ is the initial water saturation at $x_R$.

Integrating Eq. (\ref{eqn:equi}) over $[x_L, x_R]\times[0, T]$ we obtain
\begin{align}
  \label{eqn:inte}
	\int_0^T \int_{x_L}^{x_R} [ \phi \frac{\partial S_w}{\partial t} + \frac{\partial v_w }{\partial x}] \mathrm{d}x\mathrm{d} t &= \int_{x_L}^{x_R} \phi S_w(x,T) \mathrm{d}x - \int_{x_L}^{x_R} \phi S_w(x,0) \mathrm{d}x + [v_w(S_w(x_R)) - v_w((S_w(x_L))] T \nonumber\\
    & = \int_{x_L}^{x_R} \phi S_w(x,T) \mathrm{d}x - \int_{x_L}^{x_R} \phi S_w(x,0) \mathrm{d}x + [v_w(S_w^R) - q] T \nonumber\\
    &= 0,
\end{align}
in Section \ref{sec:nume} we will test the conservation property of the numerical schemes using this formula.
\subsection{Dynamic capillary pressure model}
Under equilibrium conditions, traditional models suggest the difference in phases pressure is equal to the capillary pressure. In the microscale, the capillary pressure is defined as the interfacial tension between two phases and in Darcy scale, it is usually given as a function of the wetting phase saturation:
\begin{align}\label{eqn:capi}
  p_n - p_w = p_c = P_c(S_w).
\end{align}

For non-equilibrium conditions, \cite{stauffer1978time}, \cite{hassanizadeh1990mechanics}, \cite{kalaydjian1992dynamic} proposed that the phases pressure difference can be written as a function of the capillary pressure under equilibrium condition minus the product of the saturation rate with a dynamic coefficient $\tau \text{~[Pa s]}$:
\begin{align}\label{eqn:tau}
  p_n - p_w = P_c(S_w) - \tau \frac{\partial {S}_w}{\partial t}.
\end{align}
The parameter $\tau$ is also known as damping coefficient and may still be a function of saturation \cite{joekar2010non}. Adding the dynamic capillary pressure term to Eq. (\ref{eqn:equi}) we obtain
\begin{align}
  \label{eqn:dyna}
	\phi \frac{\partial{S}_w}{\partial t} + \frac{\partial}{\partial x}\big[q f(S_w)  + \lambda_n(S_w) f(S_w) \big(\frac{\partial}{\partial x}(P_c(S_w) - \tau \frac{\partial S_w}{\partial t}) + (\rho_w - \rho_n) g\big)\big] = 0.
\end{align}
In the following we mark this model as Model 1.
% For more details the reader is referred to the mathematical theory in [15 ,11] and the references citen therein.

\subsection{Play-type capillary pressure hysteresis model }\label{sec:hyst}
\begin{figure}[!htbp]
\begin{center}
   {\includegraphics[width=2.7in] {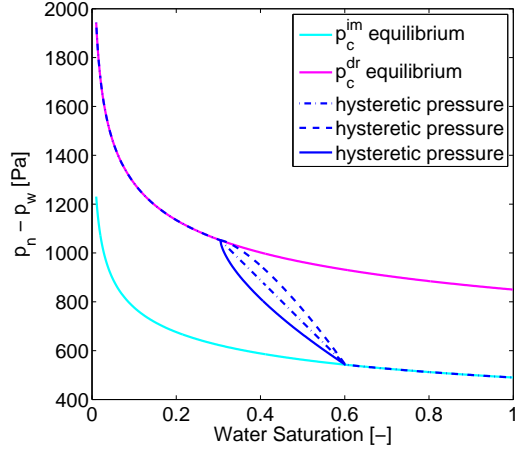}}
	 \caption{Schematic plots of capillary pressure and hysteresis loops as functions of water saturation\label{fig:hystcapi}\newline
  Cyan and magenta lines are the equilibrium imbibition and drainage capillary pressure obtained by the Brooks-Corey model in Table \ref{tab:model}; blue, dashed blue and dash-dotted blue lines illustrate an imbibition hysteresis curve, a drainage hysteresis curve and a play-type hysteresis curve. 
} 
\end{center}
\end{figure}

Many studies \cite{morrow1965capillary,jerauld1990effect} in recent decades have shown non-uniqueness in the relationship between capillary pressure and saturation, which can depend both on the history of flow displacement and the rate of change of saturation. The dependency of $p_c\text{-}S_w$ on the history of flow is known as capillary pressure hysteresis. Displacement of flow differentiates between drainage and imbibition. The process of drainage describes when the nonwetting phase displaces the wetting phase. Vice versa, imbibition describes the process when the wetting phase displaces the nonwetting phase. In general, for a given saturation $S_w$, $p_c$ can lie anywhere within the primary drainage curve $P_c^{dr}$ and the primary imbibition curve $P_c^{im}$, depending on the saturation history. Some typical plots of hysteretic capillary pressure curves are presented in Fig. \ref{fig:hystcapi}. In \cite{parlange1976capillary,beliaev2001theoretical,brokate2012numerical} different kinds of hysteresis models have been discussed for two-phase flows, in our work we adopt the play-type hysteresis model presented in \cite{brokate2012numerical}.

%As is mentioned in \cite{hassanizadeh2002dynamic}, the relationship between capillary pressure and saturation is not unique; it depends on both the history and the rate of change of saturation. Displacement of flow differentiates between drainage and imbibition. The process of drainage describes when the nonwetting phase displaces the wetting phase. Vice versa, imbibition describes the process when the wetting phase displaces the nonwetting phase.
%
%The dependency of $p_c\text{-}S_w$ on the history of flow is known as capillary pressure hysteresis. 

Assume $p_c$ depends only on $S_w$ and this relationship is described by the hysteresis operator
\begin{align}\label{eqn:hyst1}
  P_c^{hyst}: S_w(\cdot) \rightarrow p_c(\cdot).
\end{align}
Note that $P_c^{hyst}$ operates on $S_w$ as a function of time. In the drainage process, when $S_w$ decreases, $p_c$ follows the drainage pressure-saturation curve $P_c^{dr}(S_w)$. In the imbibition process, when $S_w$ increases, $p_c$ follows the imbibition pressure-saturation curve $P_c^{im}(S_w)$. In this hysteresis model, between the drainage and imbibition curves, $p_c$ and $S_w$ evolve as 
\begin{align}\label{eqn:hyst2}
  \frac{\partial p_c}{\partial t} = -\beta \frac{\partial S_w}{\partial t},
\end{align}
where $\beta$ is the opposite slope of the hysteresis curve with dimension $\text{[Pa]}$. $\beta$ is usually chosen to be very large, which means the hysteresis curve is very steep and the hysteresis pressure can move quickly from one equilibrium curve to another when the saturation direction changes. This hysteretic capillary pressure curve is illustrated in Fig. \ref{fig:hystcapi} (dashed-dotted blue line).

Since $P_c^{hyst}$ acts on the history of $S_w$, it is not possible to compute $p_c$ at a given time from $S_w$ at that time alone. Consider the system after time discretization, denote $p_c$ and $S_w$ from the previous time step as $p_c^{n-1}$ and $S_w^{n-1}$, respectively. The discrete form of Eq. (\ref{eqn:hyst2}) is
\begin{align}
	\label{eqn:hyst3}
	p_c^{n} = p_c^{n - 1} - \beta (S_w^n - S_w^{n-1}).
\end{align}
The algorithm for computing $p_c$ is as follows
\begin{enumerate}
  \item Set $p_c^{n}= p_c^{n-1} - \beta(S_w^n - S_w^{n-1})$.
  \item If $p_c^{n} < P_{c}^{im}(S_w^n)$, set $p_c^{n} = P_c^{im}(S_w^n)$.
  \item If $p_c^{n} > P_c^{dr}(S_w^n)$, set $p_c^{n} = P_c^{dr}(S_w^n)$.
\end{enumerate}
In the numerical simulations, we denote the above algorithm as $p_c^n = P_c^{hyst}(S_w^n)$.

Combine capillary pressure hysteresis with the dynamic capillary pressure (\ref{eqn:tau}) we obtain
\begin{align}\label{eqn:hystbeta}
  p_n - p_w = P_c^{hyst}(S_w) - \tau \frac{\partial S_w}{\partial t}.
\end{align}
Substituting Eq. (\ref{eqn:hystbeta}) into Eq. (\ref{eqn:equi}) we get a model with dynamic capillary pressure effect and play-type capillary pressure hysteresis, in the following this model is marked as Model 2.

\subsection{Hysteretic dynamic capillary pressure model with play-type capillary pressure hysteresis}

\cite{sakaki2010direct} conducted a series of experiments to measure values of the dynamic capillary coefficient $\tau$ for a porous medium. Their result suggests that $\tau$ is hysteretic. \cite{mirzaei2013experimental} investigated the hysteretic behaviour between $\tau$ and $S_w$. The experiments demonstrate that the value of $\tau$ for imbibition is generally larger as compared to the $\tau$ value for drainage at the same saturation. Thus it is reasonable to introduce hysteresis in the $\tau\text{-}S_w$ relationship. We assume in the imbibition process $\tau = \tau^{im}$, in the hysteresis process $\tau$ decreases from $\tau^{im}$ to $\tau^{dr}$ and at the tail the dynamic coefficient is $\tau^{dr}$. To our best knowledge, the ratio ${\tau^{dr}}/{\tau^{im}}$ has not been investigated in the literature. In this work, we follow \cite{van2007new} to show the influence of $\tau$ by using the travelling ansatz.

Eq. (\ref{eqn:dyna}) can be rewritten as
\begin{align}
		\label{eqn:fullmbl}
			\frac{\partial S_{w}}{\partial t} + \frac{\partial F(S_w) }{\partial x}= - \frac{\partial}{\partial x} [ H(S_w) \frac{\partial}{\partial x} (P_c(S_w) - \tau \frac{\partial S_{w}}{\partial t})],
\end{align}
where the flux $F(S_w)$ and the capillary induced diffusion \cite{cuesta2006non} $H(S_w)$ are given by
\begin{align}
		& F(S_w) = \frac{ 1}{\phi}f(S_w) [ v_T + \lambda_n(S_w) (\rho_w - \rho_n) g], 	& H(S_w) = \frac{1}{\phi} \lambda_n(S_w) f(S_w)\label{eqn:diff}.
\end{align}

In order to find a traveling wave solution for Eq. (\ref{eqn:fullmbl}), we introduce the new variable $\eta = x - st $. Substituting $S_w(\eta)$ into (\ref{eqn:fullmbl}) results in a third order ordinary differential equation (ODE)
\begin{equation}
\label{eqn:twfull}
    \begin{aligned}
	&-s S_w' + [F(S_w)]' = - [H(S_w) P'_c(S_w) S_w']' - s \tau [H(S_w)S_w'']', \\
  \end{aligned}
\end{equation}
where prime denotes differentiation with respect to $\eta$. This equation is to be solved subject to the boundary conditions at infinities,
\begin{align}
	\label{eqn:bc0}
	S_w(-\infty) = S_w^L, \quad S_w(\infty) = S_w^R, \quad S_w^L, S_w^R \in[0, 1].
\end{align}
Integrating Eq. (\ref{eqn:fullmbl}) over $(\eta, \infty)$ and assuming
\begin{align}
  \label{eqn:odeorder2bc}
 [H(S_w) (P'_c (S_w) S_w' - s \tau  S_w'')] (\pm\infty) = 0 ,
\end{align}
yields the second-order ODE:
\begin{equation}
	\label{eqn:odeoder2}
\left\{
\begin{aligned}
	& - s(S_w - S_w^R) + [F(S_w) - F(S_w^R)] = -H(S_w) P'_c(S_w) S_w' - s \tau H(S_w) S_w'', \\
   & S_w(-\infty) = S_w^L, \quad S_w(\infty) = S_w^R,
\end{aligned}
\right.
\end{equation}
with $s$ determined by the Rankine-Hugoniot condition
\begin{align}
	\label{eqn:rhc}
	s = \frac{F(S_w^L) - F(S_w^R)}{S_w^L - S_w^R}.
\end{align}
When gravity is included into the flux function $F(S_w)$, with different values of $v_T$, $F(S_w)$ may be non-monotone. For simplicity, we only consider $(S_w^L, S_w^R)$ pairs that satisfy $s > 0$.

Next we write Eq. (\ref{eqn:odeoder2}) as a first order system of ODEs:
\begin{equation}\label{eqn:odesorder1}
	\left\{ \begin{aligned}
		&S_w' = v, \\
		&v' = \frac{1} {s  \tau H(S_w)} \big[s(S_w- S_w^R) - [F(S_w) - F(S_w^R)] - H(S_w)P'_c(S_w) v \big].
	\end{aligned}\right.
\end{equation}
Let $S_w^\alpha$ be the unique root of the equation
\begin{align}
	F'(S_w) = \frac{F(S_w) - F(S_w^0)}{S_w- S_w^0},
\end{align}
where $S_w^0$ is the initial water saturation ahead of the wetting front.

The Jacobian of (\ref{eqn:odesorder1}) reads
\begin{equation}
	A = \left[ \begin{array}{cc}
			0 & 1\\
			\frac{s - F'(S_w) }{s \tau H(S_w) } & -\frac{H(S_w)P_c'(S_w)}{s \tau  H(S_w) }
	\end{array} \right],
\end{equation}
and has eigenvalues
\begin{align}
\label{eqn:eigenvalue}
	\lambda_{\pm} = \frac{1}{2 s \tau } [- P'_c(S_w) \pm \sqrt{(P'_c(S_w))^2 - 4 s \tau \frac{ F'(S_w) -s )}{H(S_w)}} ].
\end{align}
For the case where saturation plateau appears, consider a traveling wave connecting $S_w^L = S_w^B$ (equilibrium boundary saturation) and $S_w^R = \bar{S}_w^P$ (plateau saturation) with wave speed
\begin{align}
  s = \frac{F(S_w^L) - F(S_w^R)}{S_w^L - S_w^R}.
\end{align}
Then we can prove $(S_w^B, 0)$ is an equilibrium of system (\ref{eqn:odesorder1}). Using (\ref{eqn:eigenvalue}), if $\tau > \tau_s = \frac{H(S_w^L) P'_c(S_w^L)^2 }{4 s (F'(S_w^L) - s)}$, the equilibrium $(S_w^B, 0)$ is a spiral, oscillation will appear near $S_w^L = S_w^B$ . When only saturation overshoot appears, consider a travelling wave connecting $S_w^L = S_w^B$ and $S_w^R = S_w^0$, if $\tau > \tau_s = \frac{H(S_w^L) P'_c(S_w^L)^2 }{4 s (F'(S_w^L) - s)}$, the equilibrium $(S_w^B, 0)$ is a spiral. 

From the analysis above, we can conclude that when $\tau < \tau_s$, saturation oscillation will not appear at the drainage front, which means $\partial S_w/ \partial t \rightarrow 0^-$. Therefore, the phase pressure difference $p_n - p_w$ will tend to $P_c^{dr}$ at equilibrium.

Denoting the hysteretic dynamic coefficient as $\tau^{hyst}$, since $\tau^{hyst}$ may possibly due to the hysteresis in the retention curve \cite{sakaki2010direct}, for simplicity, we introduce $\tau^{hyst}$ as the linear interpolation between $\tau^{dr}$ and $\tau^{im}$ by utilizing the capillary pressure hysteresis,
\begin{align}
  \label{eqn:calcmtau}
  \tau^{hyst} = (\tau^{im} - \tau^{dr}) [\frac{ P_c^{hyst}(S_w) - \frac{1}{2} (P_c^{im}(S_w) + P_c^{dr} (S_w))}{P_c^{im}(S_w) - P_c^{dr}(S_w)} + \frac{1}{2}] + \tau^{dr}.
\end{align}
Substituting Eqs. (\ref{eqn:hystbeta}) and (\ref{eqn:calcmtau}) into Eq. (\ref{eqn:equi}) will result in a model with hysteretic dynamic capillary pressure and capillary pressure hysteresis. This model is marked as Model 3.

\section{Numerical scheme}
In this section we present the numerical scheme based on a reformulation of the non-equilibrium equation, the method of lines is then applied to this reformulation. Denoting $p = p_n - p_w$, Eq. (\ref{eqn:dyna}) can be rewritten as
\begin{equation}\label{eqn:line}
  \left\{
  \begin{aligned}
		&\phi \frac{P_c(S_w) - p}{\tau}  + \frac{\partial}{\partial x} [q f(S_w) + \lambda_n(S_w) f(S_w) (\frac{\partial p}{\partial x} + (\rho_w - \rho_n) g)] = 0, \\
 &\frac{\partial S_w}{\partial t} = \frac{P_c(S_w) - p}{\tau}.
  \end{aligned}
  \right.
\end{equation}
Since Model 2 and Model 3 are incorporated with the capillary pressure hysteresis, we replace $P_c(S_w)$ in Eq. (\ref{eqn:line}) by $P_c^{hyst}(S_w)$ when solving these two models and replace $\tau$ by $\tau^{hyst}$ when solving Model 3.

\subsection{Castillo-Grone mimetic operators}\label{sec:cgm}
To discretize Eq. (\ref{eqn:line}) in the space direction we adopt the mimetic finite difference method which satisfies the discrete version of continuum conservation law. \cite{castillo2003matrix} developed a set of mimetic operators knows as Castillo-Grone's mimetic (CGM) operators. CGM operators have been used in many fields, such as seismic studies \cite{rojas2008modelling}, electrodynamics \cite{runyan2011novel} and image processing \cite{bazan2011mimetic}. Numerical results in these fields validate the high efficiency and reliability of the CGM operators.

The main features of CGM operators are that they preserve symmetry properties of the continuum and have overall high order accuracy. The CGM operators can be implemented as efficient as the standard finite difference schemes. Here, we briefly describe the CGM operators in one dimension as applied in this work. The CGM 2-D operators can be obtained by the Kronecker products of block matrices. For more details, see \cite{castillo2013mimetic} and references therein.

%%%%%%%%%%%%%%%%%%%%%%%%%%%%%%%%%%%%%%%%%%%%%%%%%
In the one-dimensional situation the Green-Gauss-Stokes theorem reads
\begin{align}
  \label{eqn:intebyparts}
	\int_{x_L}^{x_R} (f \frac{\partial v}{\partial x} + \frac{\partial f}{\partial x} v) \mathrm{d}x = f(x_R) v(x_R) - f(x_L) v(x_L),
\end{align}
where $f$ and $v$ are two smooth real-valued functions defined in interval $\Omega = [x_L, x_R]$. Let $L = x_R - x_L$, and the step size $\Delta x = L / N$, then $\Omega$ can be partitioned into $N$ equal sized cells $[x_i, x_{i+1}]$, where $0 \leq i \leq N-1$. The cell centers can be indexed as $x_{i+1/2} = \frac{1}{2} (x_i + x_{i+1})$ for $0 \leq i \leq N-1$. The cell nodes $x_i$ and cell centers $x_{i+1/2}$ build up the uniform staggered grid. We discretize $v$ at the cell nodes:
\begin{align}
  \bar{v} = [v(x_0), v(x_1), \cdots, v(x_{N-1}), v(x_N)]^T,
\end{align}
and $f$ at the cell centers and the boundary nodes:
\begin{align}
  \hat{f} = [f(x_0), f(x_{1/2}), \cdots, f(x_{N-1/2}), f(x_N)]^T.
\end{align}
Let $\hat{D}_{(N+2)\times (N+1)}$ denote the CGM divergence operator and $G_{(N+1)\times(N+2)}$ denote the CGM gradient operator, then the discrete version of the conservation law  (\ref{eqn:intebyparts}) reads
\begin{align}\label{eqn:disclaw}
  <\hat{D} \bar{v}, \hat{f}>_Q + <\bar{v}, G \hat{f}>_P = <\hat{B} \bar{v}, \hat{f}>_I,
\end{align}
where $<x, y>_A = y^T A x$ is the inner product, $I$ is the $(N+2) \times (N+2)$ identity matrix, $Q$ and $P$ are the weight matrices for $\hat{D}$ and $G$ respectively. The matrix $\hat{B} = Q \hat{D} + G^T P$ embodies the global conservation requirement for the discrete conservation law and is called the boundary operator. \cite{castillo2005linear} presented the second-order divergence mimetic operator as
\begin{equation}
  \hat{D} = \left[ \begin{array}{ccc}
          0 & \dots  & 0 \\
            & D     & \\
          0 &   \dots & 0
          \end{array} \right] \in R^{(N+2)\times (N+1)},
\end{equation}
where
\begin{equation}
   D = \frac{1}{\Delta x}\left[ \begin{array}{cccccc}
                  -1 & 1 & 0 & \cdots & \cdots & 0 \\
                   0  & -1 & 1 & 0 & \cdots & \vdots\\
                   0  &  \ddots  & \ddots &\ddots & \ddots& \vdots \\
                   \vdots  & \ddots& 0 & -1 & 1 & 0\\
                  0   &\cdots & \cdots& 0& -1 & 1
                  \end{array}
        \right] \in R^{N \times (N+1)}.
\end{equation}
For this second-order mimetic divergence matrix ${\hat{D}}$, the weights matrix $Q$ is the $(N+2) \times (N+2) $ identity matrix. The second order CGM gradient operator reads as
\begin{equation}
 G = \frac{1}{\Delta x} \left[ \begin{array}{cccccc}
                  -8/3 & 3 & -1/3 & 0 & \cdots & 0\\
                    0 & -1 & 1 &0 &  \cdots & \vdots\\
                    0 & \ddots  & \ddots &\ddots & \ddots& \vdots\\
                    \vdots &\ddots &0 & -1 & 1 & 0\\
                     0 &\cdots & 0 & 1/3& -3 & 8/3
                  \end{array}
        \right] \in R^{(N+1) \times (N+2)},
\end{equation}
with weight matrix
\begin{equation}
  P = \left[ \begin{array}{ccccccc}
      3/8 & 0 & \cdots & \cdots & \cdots & \cdots & 0  \\
      0 & 9/8 & \ddots & & & & \vdots \\
      \vdots & \ddots & 1 & \ddots & & & \vdots \\
      \vdots &         & \ddots & \ddots & \ddots& & \vdots \\
      \vdots &        &        &  \ddots  & 1 & \ddots & \vdots \\
      \vdots &        &        &      & \ddots & 9/8 & 0 \\
      0 & \cdots & \cdots &\cdots & \cdots & 0 & 3/8\\
      \end{array}\right] \in R^{(N+1) \times (N+1)}.
\end{equation}

Applying the discrete conservation law (\ref{eqn:disclaw}) with matrices $D$ and $G$, the boundary operator $\hat{B}$ can be written as
\begin{equation}
  \hat{B} = \left[ \begin{array}{cccccccc}
  -1 & 0 & 0 & \cdots & \cdots & 0 & 0 & 0 \\
  1/8& -1/8 & 0 & \cdots & \cdots &0  & 0 &0 \\
  -1/8 & 1/8 & 0 &\cdots & \cdots & 0& 0 & 0\\
  0 & 0 & 0 & \ddots & & 0 & 0 & 0 \\
  \vdots & \vdots & \vdots & \ddots & \ddots & \vdots & \vdots &\vdots \\
  0 & 0 & 0 & & \ddots & 0 & 0 & 0 \\
  0 & 0 & 0 & \cdots & \cdots & 0 & -1/8 & 1/8 \\
  0 & 0 & 0 & \cdots & \cdots & 0 & 1/8 & -1/8 \\
  0 & 0 & 0 & \cdots & \cdots & 0 & 0 & 1
  \end{array} \right] \in R^{(N+2)\times(N+1)}.
\end{equation}

%%%%%%%%%%%%%%%%%%%%%%%%%%%%%%%%%%%%%%%%%%%%%%%%%%
\subsection{Mimetic discretizations for the two-phase flow equations}
Using the uniform staggered grid presented in Section \ref{sec:cgm}, we discretize $S_w$ and $p$ at the centers as $\bar{S}_w = [S_{w0}, S_{w\frac{1}{2}}, \cdots, S_{w\frac{N-1}{2}}, S_{wN}]$ and $\bar{p} = [p_{0}, p_{\frac{1}{2}}, \cdots, p_{\frac{N-1}{2}}, p_{N}]$. Then we use linear interpolation to get $S_w$ at the nodes
\begin{align}\label{eqn:lineinte}
  \hat{S}_{wi} = \frac{1}{2}(S_{wi-\frac{1}{2}} + S_{w i+ \frac{1}{2}}), \quad i = 1, \cdots, N-1,
\end{align}
At the boundary $\hat{S}_{w0} = S_{w0}$, $\hat{S}_{wN} = S_{wN}$.

Introducing coefficients matrix $K\in R^{(N+1)\times(N+1)}$ with zero non-diagonal elements, the diagonal elements are given by
\begin{align}
  K_{ii} &=   \lambda_n(\hat{S}_{wi-1}) f(\hat{S}_{wi-1}), \quad i= 1, 2 \cdots, N+1.
\end{align}

In the numerical simulations, since fully implicit time discretizations allowing large time steps are preferred for solving long-time scale problems, thus we apply the implicit trapezoidal integration to (\ref{eqn:line}) in the time direction. Discretizing (\ref{eqn:line}) and boundary condition (\ref{eqn:bc}) with the CGM operators in the space direction, we obtain
\begin{equation}
  \label{eqn:fulldisc}
\begin{aligned}
	(\phi \hat{I} - \tau \hat{I} \hat{D} K  G + K_{11} B G + A) \bar{p}^{n+1}& = \phi \hat{I} {P}_c(\bar{S}_w^{n+1}) \\
&	+ \tau \hat{D} [ q f(\hat{S}_w^{n+1}) + \lambda_n(\hat{S}_w^{n+1}) f(\hat{S}_w^{n+1}) (\rho_w - \rho_n) g] + \bar{b}(\bar{S}_w^{n+1}),
\end{aligned} 
\end{equation}
and
\begin{equation}
\label{eqn:disctime}
\begin{aligned}
	&\bar{S}^{n+1}_w  = \bar{S}_w^n + \frac{\Delta t }{2} [\frac{P_c(\bar{S}_w^n) - \bar{p}^n)}{\tau} + \frac{P_c(\bar{S}_w^{n+1}) - \bar{p}^{n+1}}{\tau}].
\end{aligned}
\end{equation}
In (\ref{eqn:fulldisc}), $\hat{I}$ is a $(N+2)\times (N+2)$-dimensional matrix with $\hat{I}_{11} = 0, \hat{I}_{N+2,N+2} = 0, \hat{I}_{ii} = 1, \text{for~} i = 2, 3, \cdots, N+1$, the matrices $A$, $B$ and vector $b(\bar{S}_w^{n+1})$ represent the boundary conditions. As a result of the flux boundary condition at $x_L$ and the Dirichlet boundary condition at $x_R$, the only non-zero element in $A$ is $A_{N+2,N+2} = 1$, and $B$ differs from $\hat{B}$ in the last three rows,
\begin{equation}
  B = \left[ \begin{array}{cccccccc}
  -1 & 0 & 0 & \cdots & \cdots & 0 & 0 & 0 \\
  1/8& -1/8 & 0 & \cdots & \cdots &0  & 0 &0 \\
  -1/8 & 1/8 & 0 &\cdots & \cdots & 0& 0 & 0\\
  0 & 0 & 0 & \ddots & & 0 & 0 & 0 \\
  \vdots & \vdots & \vdots & \ddots & \ddots & \vdots & \vdots &\vdots \\
  0 & 0 & 0 & & \ddots & 0 & 0 & 0 \\
  0 & 0 & 0 & \cdots & \cdots & 0 & 0 & 0 \\
  0 & 0 & 0 & \cdots & \cdots & 0 & 0 & 0 \\
  0 & 0 & 0 & \cdots & \cdots & 0 & 0 & 0
  \end{array} \right] \in R^{(N+2)\times(N+1)},
\end{equation}
The $(N+2)\times 1$ column vector $\bar{b}(\bar{S}_w^{n+1})$ reads
\begin{equation}
	{b}(\bar{S}_w^{n+1}) = \left[ \begin{array}{c}
			q  - q f(S_{w0}^{n+1}) - \lambda_n(S_{w0}^{n+1}) f(S_{w0}^{n+1}) (\rho_w - \rho_n) g \\
    0 \\
    \vdots \\
    0 \\
    P_c(S_w^{R}) \\
     \end{array}\right].
\end{equation}

In order to solve (\ref{eqn:fulldisc}) and (\ref{eqn:disctime}) we apply the iteration method. By introducing the superscript $l$ as an iteration counter, the algorithm for each time step is as follows:
\begin{enumerate}
	\item Set $\bar{S}_w^{n+1,0} = \bar{S}_w^n$, $\bar{p}^{n+1,0} = \bar{p}^n$, $P_c(\bar{S}_w^{n+1, 0}) = P_c(\bar{S}_w^{n})$ (or $P_c^{hyst}(\bar{S}_w^{n+1,0})=P_c^{hyst}(\bar{S}_w^n$)), $l = 0$.
	\item Update $\bar{S}_w^{n+1,l+1} = \bar{S}_w^n + \frac{\Delta t }{2} [\frac{P_c(\bar{S}_w^n) - p^n)}{\tau} + \frac{P_c(\bar{S}_w^{n+1,l}) - \bar{p}^{n+1,l}}{\tau}]$, solve (\ref{eqn:fulldisc}) for $\bar{p}^{n+1,l+1}$ and update $P_c(\bar{S}_w^{n+1,l+1})$ (or $P_c^{hyst}(\bar{S}_w^{n+1,l+1})$).
	\item $l = l+1$.
	\item Repeat steps 2 and 3, until $|\bar{S}_w^{n+1,l+1} - \bar{S}_w^{n+1, l}| < \mathrm{tol}$.
	\item Set $\bar{S}_w^{n+1} = \bar{S}_w^{n+1,l+1}$, $p^{n+1} = p^{n+1,l+1}$.
\end{enumerate}
In step 2, Eq. (\ref{eqn:fulldisc}) is a linear system in $\bar{p}^{n+1}$ which can be solved by a linear solver, in this work we adopt the build-in backslash operator of Matlab \cite{matlabr2014a} to solve $\bar{p}^{n+1}$. 

\textit{Remark:} The application of MFD to partial differential equations constitutes an active filed of research \cite{lipnikov2014mimetic}. Formal analysis of MFD for the Richards equation can be achieved by combining the convergence results in \cite{brezzi2005convergence}, with the equivalence between MFD and multipoint flux approximation (MPFA) established in \cite{stephansen2012convergence} and the convergence proof of MPFA for Richards equation in \cite{klausen2008convergence}. However, difficulties arise when establishing convergence for (\ref{eqn:line}), because of the nonlinearity and the reformulation. In Section \ref{sec:nume}, we present numerical results to demonstrate the convergence of the method.

\section{Numerical experiments}\label{sec:nume}
\cite{dicarlo2004experimental,dicarlo2007capillary} presented snapshots of the saturation and capillary pressure profiles for different fluxes in initially dry $20/30$ sand. The physical parameters of the $20/30$ sand \cite{dicarlo2004experimental,schroth1996characterization} as well as the constants and Brooks-Corey models \cite{brooks1966properties} are listed in Table \ref{tab:2030sand} and Table \ref{tab:model}. DiCarlo observed that for the highest ($q = 2.0\text{e-3} \mathrm{~[m~s^{-1}]}$) and lowest ($q = 1.32\text{e-07}\mathrm{~[m~s^{-1}]}$) fluxes, the saturation profiles are monotonic with distance and no saturation overshoot is observed, while all of the intermediate fluxes exhibit saturation overshoots. In this section we will study the numerical behaviours of the three models presented in Section \ref{sec:models}.

\begin{table}
  \caption{\label{tab:2030sand}Physical parameters for $20/30$ sand.}\footnotetext{ The residual saturations in \cite{dicarlo2004experimental,dicarlo2008nonmonotonic} are different, so we simply set $S_{wr} = 0$.}
  \center
  \begin{tabular}{|cccccc|ccc|}\hline
  {}& {} & {} & \multicolumn{3}{c|}{Drainage}&\multicolumn{3}{c|}{Imbibition}\\\cline{4-9}
{Sand} & $\kappa~\mathrm{[m~s^{-1}]}$ & $\phi$~[-] & $S_{wr}$~[-] & $\lambda$~[-]& $p_d ~\text{[Pa]}$ & $S_{wr}$~[-] & $\lambda$~[-] & $p_d ~\text{[Pa]}$\\\hline
    $20/30$ & $2.5\text{e-03}$ & $0.35$ & $0$ & $5.57$ & $850$ & $0$ & $5$ & $490$ \\\hline
  \end{tabular}
\end{table}

\begin{table}
  \caption{\label{tab:model}Constants and Brooks-Corey models.}
  \center
  \begin{tabular}{|c|cc|}\hline
  Density $\mathrm{[kg~m^{-3}]}$ & $\rho_w = 998.21$ & $\rho_n = 1.2754$  \\
  Viscosity $\mathrm{[kg~m^{-1} s^{-1}]}$ & $\mu_w = 1.002\text{e-03}$ & $\mu_n = 1.82\text{e-05}$  \\
  Mobility $\mathrm{[m~s~kg^{-1}]}$ & $\lambda_w = \frac{K k_{rw}}{\mu_w}$ & $\lambda_n = \frac{K k_{rn}}{\mu_n}$ \\
  Constants & $g = 9.81 ~\mathrm{[m~s^{-2}]}$  & $K = \frac{\kappa \mu_w}{\rho_w g}$ $~\mathrm{[m^2]}$\\\hline
   & Capillary pressure  &{Relative permeability } \\\hline
   & $S_e = \frac{S_w - S_{wr}}{1 - S_{wr}}$ &  $k_{rw} = S_e^{\frac{2+ 3 \lambda}{\lambda}}$ \ \\
  \raisebox{1.6ex}[0pt]{Brooks-Corey model} & {
  $p_c = p_d S_e^{-\frac{1}{\lambda}},~~\mathrm{for}~p_c > p_d$}& $k_{rn} = (1-S_e)^2 (1-S_e^{\frac{2+\lambda}{\lambda}})$ \\\hline
  \end{tabular}
\end{table}

In Eq. (\ref{eqn:equi}) when $S_w = 0$ or $S_w = 1$ the equation is degenerate. From Fig. 1 in \cite{dicarlo2007capillary} we get the initial capillary pressure $p_c^{0} \approx 1600 \text{[Pa]}$, using the Brooks-Corey capillary pressure model in Table \ref{tab:model} we can find for the imbibition process, when water saturation $S_w = 0.003$, the Brooks-Corey capillary pressure $p_{c}(S_w) = 1566 \text{[Pa]}$. So in the numerical simulations, we set $S^{R}_w = 0.003$ and $S^{max}_w = 1 - 1.0\text{e-03}$. The initial saturation is given by
\begin{align}\label{eqn:initcond}
  S_w(x, 0) = S_{w}^{R} + (S_{w}^L - S_{w}^{R}) (1 - \tanh(200x)),
\end{align}
where $S_w^L = 0.025$, the initial phases pressure difference is $p_n - p_w = P_c(S_w(x,0))$. The reason we set $S_w^L > S_w^R$ is that, when $S_w^L = 0.003$, in Eq. (\ref{eqn:disctime}) we have $\lambda_n(0.003)f(0.003) \approx 6.7\text{e-16} $, then the boundary saturation obtained by solving (\ref{eqn:line}) will exceed 1; when $S_w$ is big enough, for example $S_w^L = 0.025$, we have $\lambda_n(0.025)f(0.025) \approx 9.1\text{e-13}$, the boundary saturation will not exceed 1, see Fig. \ref{fig:case4_bc_u_t}. For the numerical simulations, $S_w^L$ is small enough and will not influence the behaviour of the models. Before we carry out the numerical simulations, the parameters $\beta$ and $\tau^{dr}$ appear in Eq. (\ref{eqn:hyst2}) and Eq. (\ref{eqn:calcmtau}) have to be decided. Here we choose $\beta =1.0\textrm{e05}$, the ratio $\tau^{dr}/\tau^{im}$ is set to be $0.2$. 

First, we test the accuracy of schemes (\ref{eqn:fulldisc}) and (\ref{eqn:disctime}). Since exact solutions of Eq. (\ref{eqn:dyna}) are not known, the numerical solutions on fine grids are taken as reference solutions. For space and time accuracy tests, the reference grids are $N = 1024, \Delta t = 0.01$ and $N = 512, \Delta t= 0.001$, respectively. Setting $S_w^L = 0.45, S_w^{R} = 0.3, q = 1.32\text{e-04}, \tau= 4.0\text{e03}$, fixing time or space step size, the $L^2$ errors and orders are obtained in Tables \ref{tab:spacconv} and \ref{tab:timeconv}. Table \ref{tab:spacconv} shows that in the space direction, schemes (\ref{eqn:fulldisc}) and (\ref{eqn:disctime}) are second order when applying to different models, and the $L^2$ errors of the three models are consistent. However, as a result of the hysteresis effects, in the time direction the $L^2$ errors increase when schemes (\ref{eqn:fulldisc}) and (\ref{eqn:disctime}) are applied to Model 2 and Model 3, also the convergence rates drop from two to about one.

\begin{table}
  \caption{\label{tab:spacconv}Space accuracy test of schemes (\ref{eqn:fulldisc}) and (\ref{eqn:disctime}) at $T = 100$ ($\Delta t = 0.01$).}
  \center
  \begin{tabular}{|c|cc|cc|cc|}\hline
{}&\multicolumn{2}{c|}{Model 1}&\multicolumn{2}{c|}{Model 2} &\multicolumn{2}{c|}{Model 3} \\\cline{2-7}
\raisebox{1.6ex}[0pt]{$
N$}&$L^{2}$ error&$L^2$ order &$L^2$ error &$L^2$ order & $L^2$ error & $L^2$ order \\ \hline
64            & 4.1726e-04 & $--$
& 3.6657e-04 & $--$ & 3.6661e-04 & $--$\\
128           & 1.0224e-04 & 2.0289 & 7.9465-05 & 2.2057
& 7.9475e-05 & 2.2057 \\
256           & 2.4289e-05 & 2.0736 & 1.7996e-05 & 2.1426
& 1.7998e-05 & 2.1427\\
512           & 4.8183e-06 & 2.3337 & 4.0298e-06 & 2.1589 
& 4.0300e-06 &2.1590 
\\ \hline
\end{tabular}
\end{table}
\begin{table}
  \caption{\label{tab:timeconv}Time accuracy test of schemes (\ref{eqn:fulldisc}) and (\ref{eqn:disctime}) at $T = 100$ ($N = 256$)}
  \center
  \begin{tabular}{|c|cc|cc|cc|}\hline
{}&\multicolumn{2}{c|}{Model 1}&\multicolumn{2}{c|}{Model 2} &\multicolumn{2}{c|}{Model 3} \\\cline{2-7}
\raisebox{1.6ex}[0pt]{$
\Delta t$}&$L^{2}$ error &$L^2$ order &$L^{2}$ error &$L^2$ order
& $L^2$ error & $L^2$ order \\\hline
0.016         & 1.0078e-08 & $--$ & 5.5162e-05 & $--$ 
& 6.0156e-05 & $--$  \\
0.008         & 2.5009e-09 & 2.0107 & 2.5972e-05 & 1.0867
& 2.8378e-05 & 1.0839  \\
0.004         & 5.9543e-10 & 2.0704 & 1.1182e-05 & 1.2158
& 1.2230e-05 & 1.2144 \\
0.002         & 1.1908e-11 & 2.3219 & 3.7358e-06 & 1.5816
& 4.0881e-06 & 1.5809 \\ \hline
\end{tabular}
\end{table}

\begin{figure}[!htbp]
\begin{center}
   \subfigure[\label{fig:case4_bc_1}Model 1, $t=425~\text{[s]}$]
   {\includegraphics[width=2.5in] {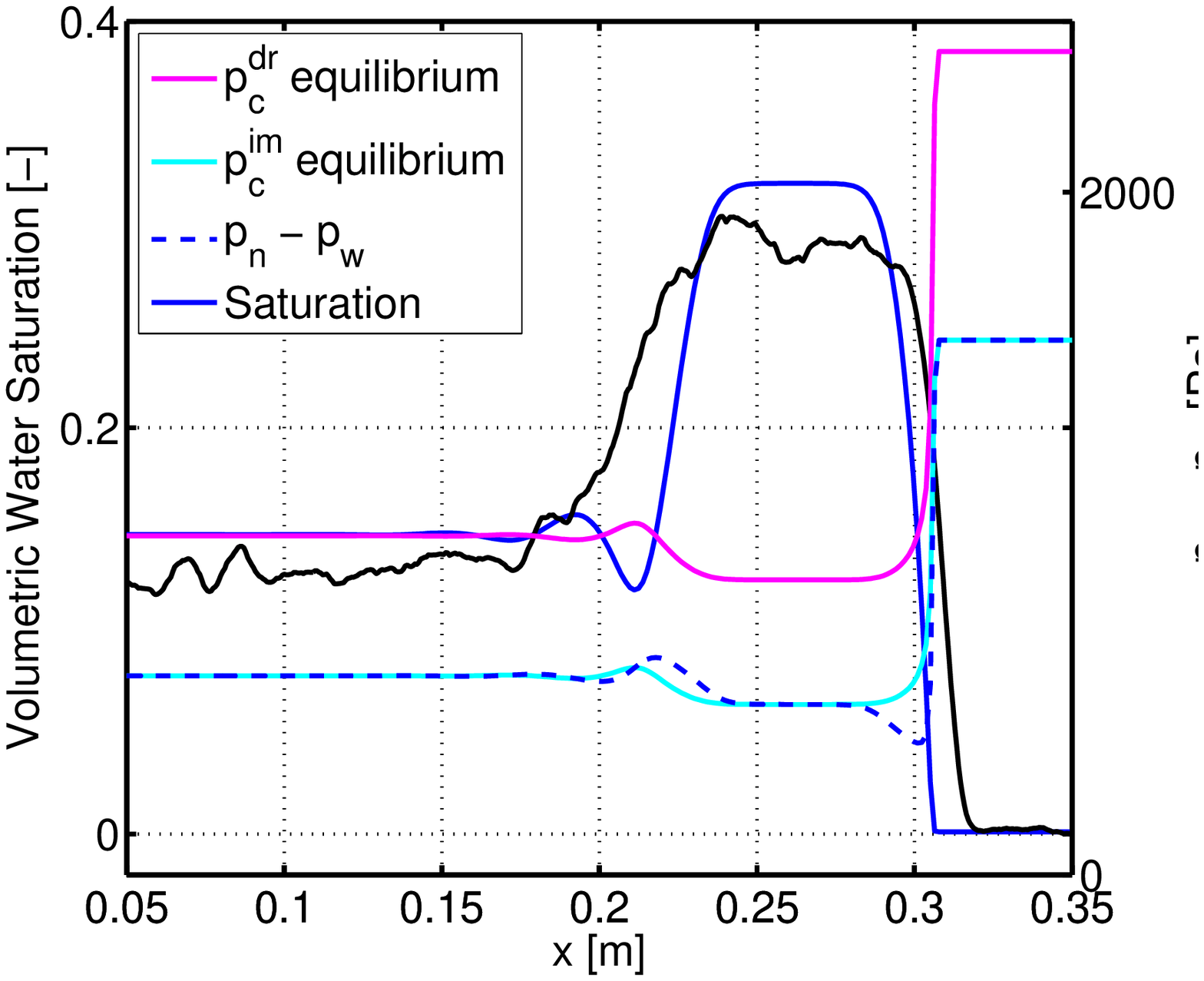}}\quad \quad
    \subfigure[\label{fig:case4_bc_2}Model 2, $t = 460~\text{[s]}$]
   {\includegraphics[width=2.5in] {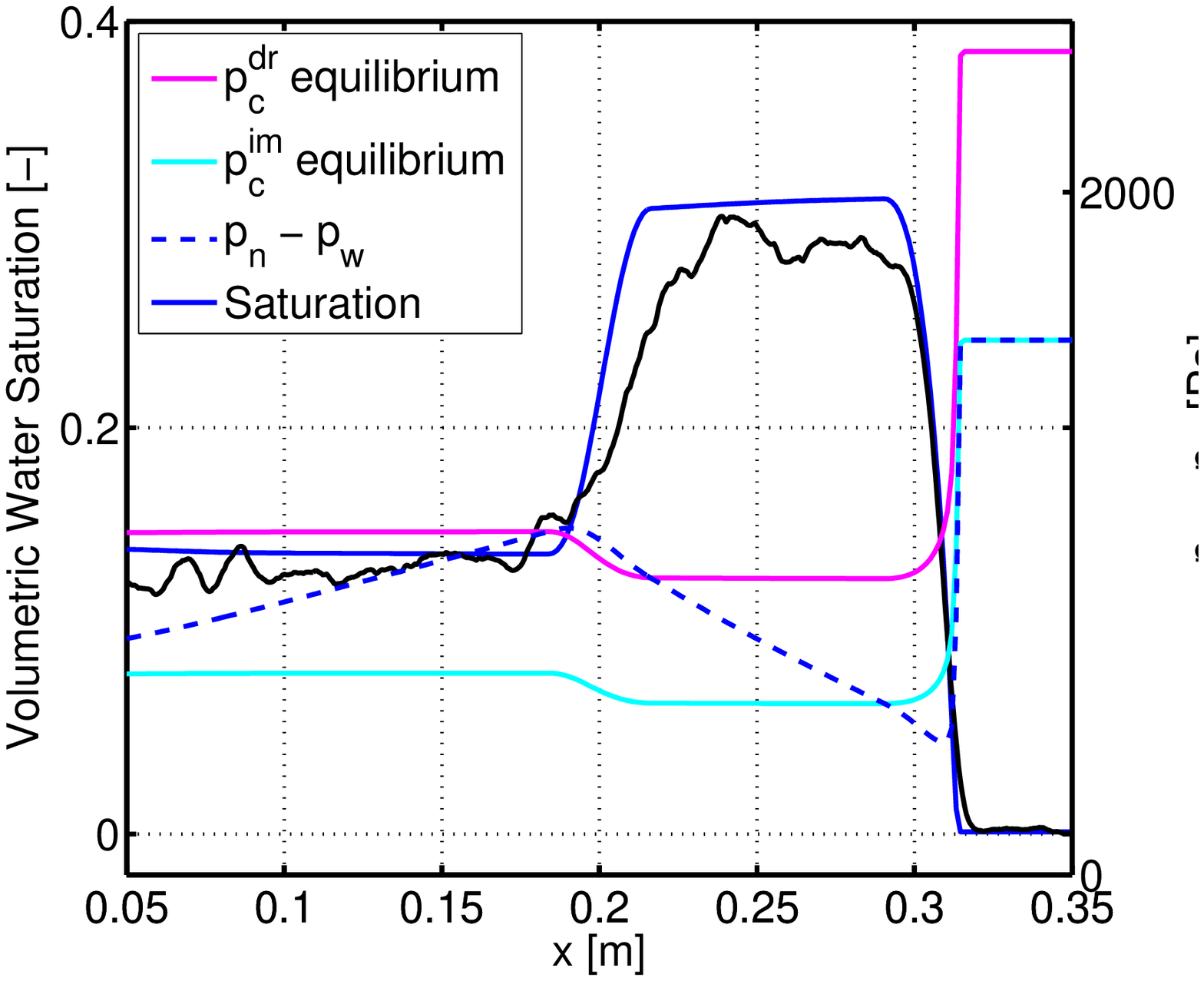}}
   \subfigure[\label{fig:case4_bc_3}Model 3, $t=460~\text{[s]}$]
   {\includegraphics[width=2.5in] {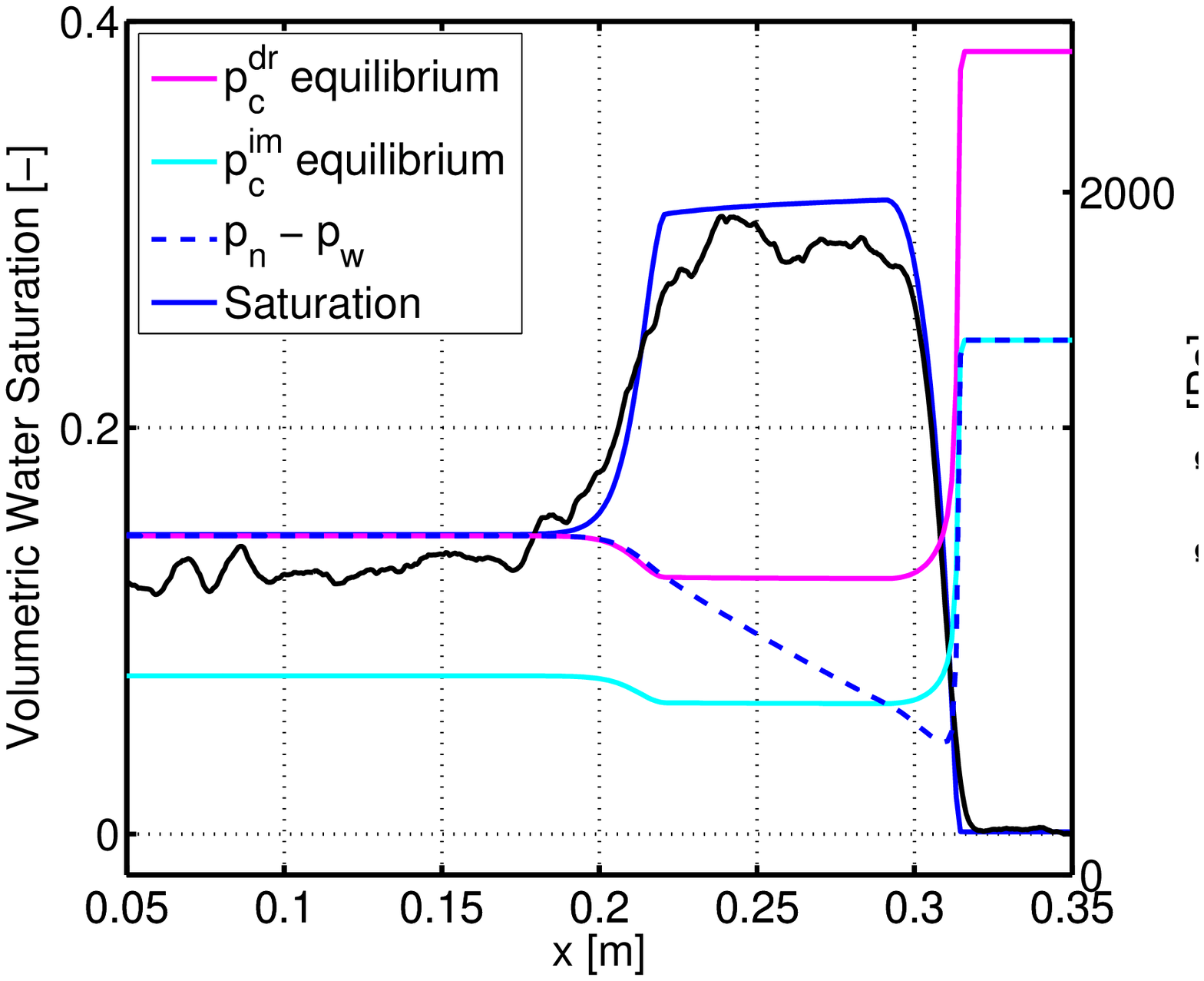}}\quad \quad
   \subfigure[Phases pressure difference-saturation, $t = 460~\text{[s]}$]
   {\includegraphics[width=2.5in] {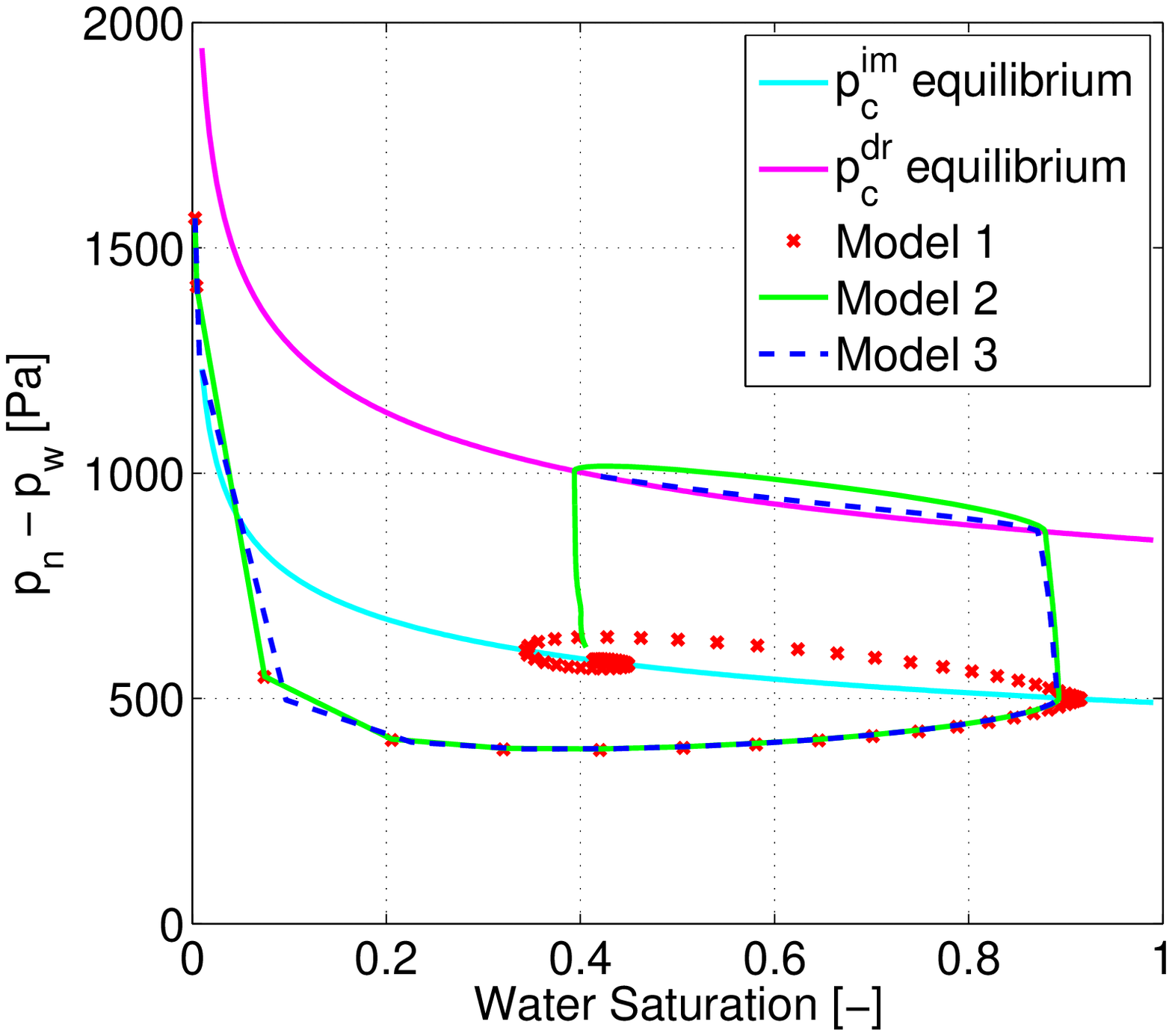}\label{fig:case4_bc_p_u}}
	 \caption{Numerical results of Model 1, 2, 3 with $q = 1.32\text{e-04}, \tau = 4.0\text{e03}$, the volumetric saturation is defined by $\phi S_w$. \newline
  (a), (b), (c) black, blue, dashed blue, cyan and magenta lines denote experimental saturation, numerical saturation, numerical capillary pressure, equilibrium imbibition and drainage pressure respectively.\newline
  (d) Cyan and magenta lines are equilibrium imbibition and drainage capillary pressure curves obtained by Brooks-Corey model. Red cross, green line and dashed blue line denote phases pressure difference-saturation relationship obtained by Model 1, 2 and 3 respectively.} \label{fig:case4_u_p_all}
\end{center}
\end{figure}

Setting $S_w^L = 0.025, S_w^{R} = 0.003, q = 1.32\text{e-04}, \tau= 4.0\text{e03}$, the saturation and pressure profiles obtained by Model 1, 2 and 3 are presented in Fig. \ref{fig:case4_u_p_all}. The imbibition fronts obtained by the three models are similar, but the saturations and pressures at the plateaus and behind drainage fronts are different. For Model 1, the value of the plateau saturation is constant, behind the drainage front, oscillations appear and the phases pressure difference follows the imbibition capillary pressure. For Model 2, the plateau saturations decrease a little from the imbibition front to the drainage front, behind the drainage front there is a slight oscillation in the saturation profile and the phase pressure difference moves to the imbibition capillary pressure. For Model 3, because of the hysteresis in the capillary pressure and the dynamic coefficient, no oscillation appears in the saturation profile. As a result, the pressure keeps constant and follows the drainage capillary pressure. The phases pressure difference-saturation curves are presented in Fig. \ref{fig:case4_bc_p_u}. The result obtained by Model 3 shows similar behaviour as the measured data in Fig. 6 in \cite{dicarlo2007capillary}.

\begin{figure}[!htbp]
\begin{center}
 {\includegraphics[width=2.7in] {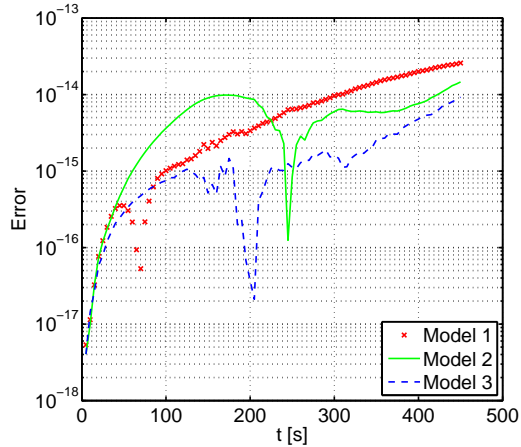}}
\caption{Errors of Eq. (\ref{eqn:inte}) using Models 1, 2 and 3  with $q = 1.32\text{e-04}, \tau = 4.0\text{e03}, N = 256$.\label{fig:case4_bc_errorinmass}}
\end{center}
\end{figure}

\begin{figure}[!htbp]
\begin{center}
 \subfigure[Saturation at left boundary over time\label{fig:case4_bc_u_t}]
   {\includegraphics[width=2.5in] {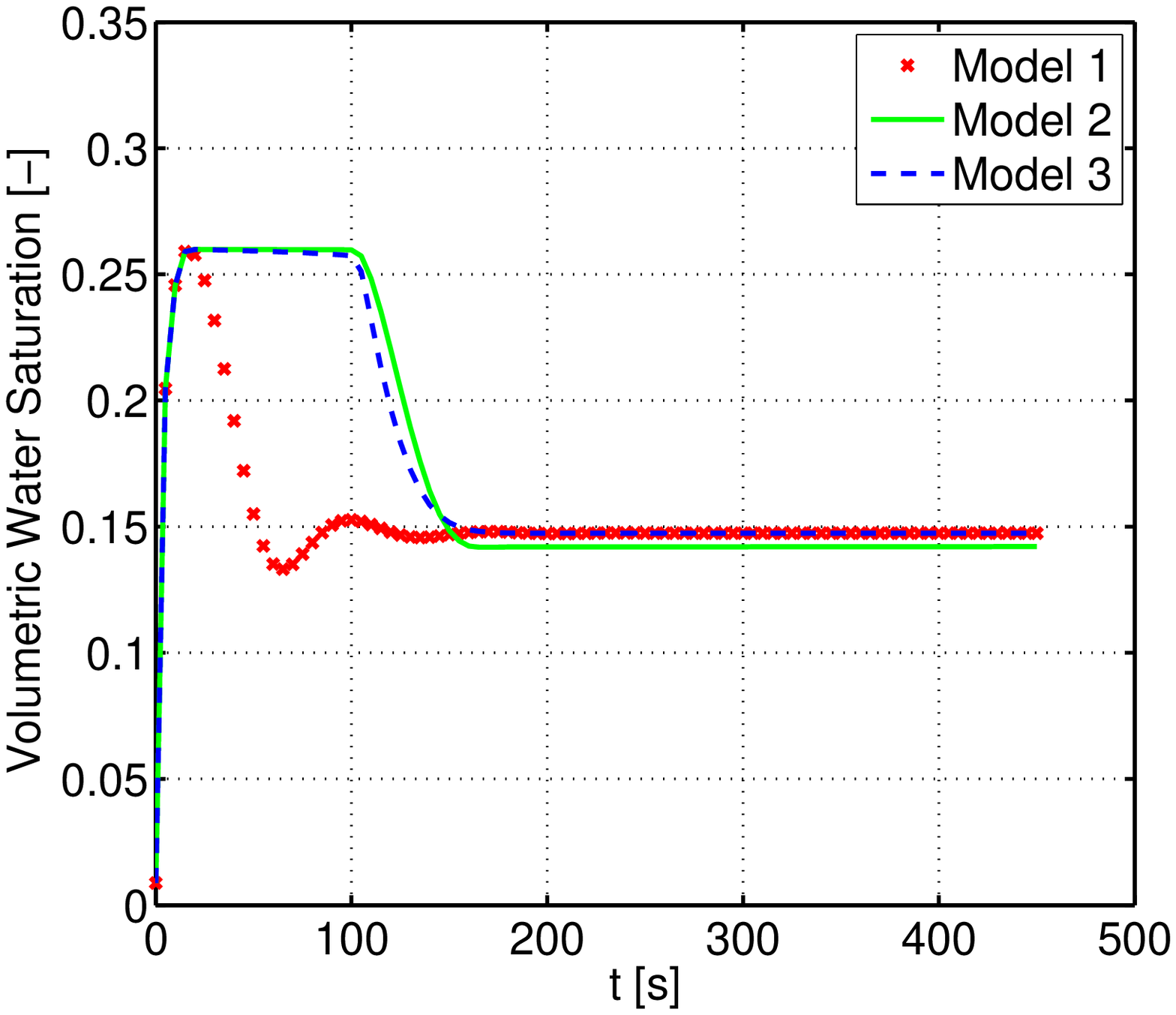}}\quad \quad
\subfigure[Phases pressure difference at left boundary over time \label{fig:case4_bc_p_t}]
 {\includegraphics[width=2.5in] {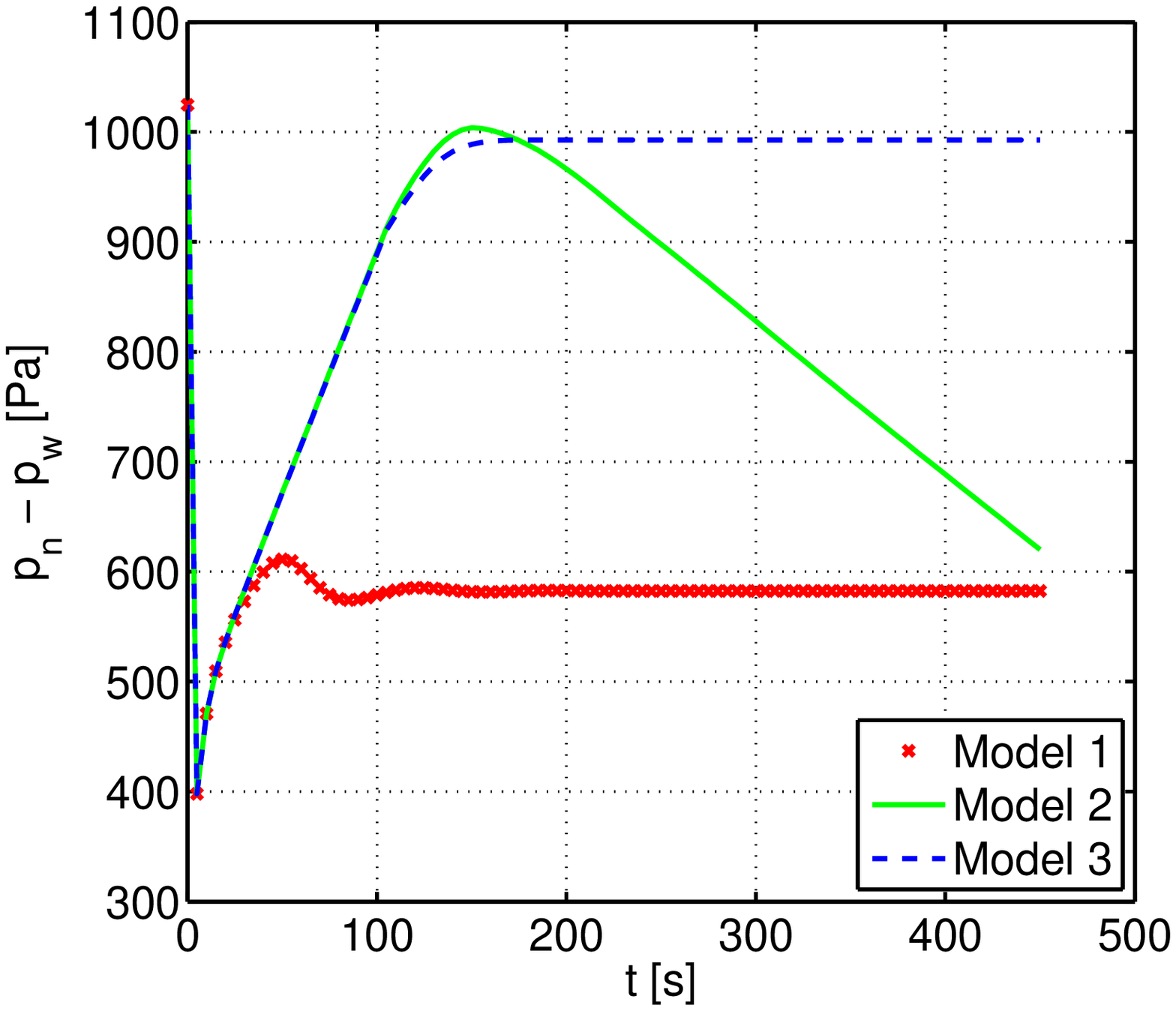}}
\caption{Saturation and phases pressure difference at left boundary, with $q = 1.32\text{e-04}, \tau = 4.0\mathrm{e03}$. \label{fig:case4_bc_up_t}}
\end{center}
\end{figure}

Fig. \ref{fig:case4_bc_errorinmass} shows that schemes (\ref{eqn:fulldisc}) and (\ref{eqn:disctime}) preserve Eq. (\ref{eqn:inte}) with high accuracy for all three models. The evolutions of saturation and pressure at the left boundary are presented in Fig. \ref{fig:case4_bc_up_t}. Fig. \ref{fig:case4_bc_u_t} shows the saturation obtained by Model 1 drops to the tail saturation after it reaches a high value, while the saturations of Model 2 and Model 3 keep the high values for a while. This phenomenon can be explained by the capillary pressure hysteresis in Model 2 and Model 3 as is shown in Fig. \ref{fig:case4_bc_p_t}. From $t = 0$ to $t \approx 100$ the hysteretic pressures in Model 2 and 3 increase, the pressure gradients keep the saturations stay at high values, when phases pressure differences reach the equilibrium drainage pressure, the pressure gradients vanish and the saturations move to the asymptotic tail values. The pressure curves obtained by Model 2 and Model 3 are also different. In Model 3, when pressure moves to the equilibrium drainage pressure, the hysteretic dynamic coefficient $\tau^{hyst}$ also decreases from $\tau^{im}$ to $\tau^{dr}$, thus keeps the saturation constant at the tail and the pressure stays at the equilibrium drainage pressure. The evolutions of the boundary saturation and pressure of Model 3 have good agreement with the observed and calculated profiles in \cite{shiozawa2004unexpected}.

Before we apply different fluxes $q$ to the three models, we have to know the end times for the simulations. In \cite{dicarlo2004experimental}, the end times for the experiments are not given, but the inner diameter of the tube is given as $d = 1.27\text{e-02~[m]}$, then the total volume of water injected into each tube can be calculated by $Volume = \phi \pi (\frac{d}{2})^2 \sum_{k = 0}^{N-1} (x_{k+1} - x_{k}) \frac{S_w(x_k) +  S_w(x_{k+1})}{2}$, where $x_k$ is the sample point in \cite{dicarlo2004experimental}. Thus we can calculate the end times using $T_{end} = \frac{Volume}{q \pi (d/2)^2} \text{~[s]}$.
\begin{table}
  \caption{\label{tab:endtime}Parameters for different fluxes.}
  \center
  \begin{tabular}{|c|cccccc|}\hline
  Volume of water $\mathrm{[m^{-3}]} $ & $1.26\text{e-05}$ & $1.17\text{e-05}$ & $6.85\text{e-06}$ & $4.56\text{e-06}$ & $2.06\text{e-06}$ & $2.18\text{e-06}$ \\ \hline
{$q \mathrm{~[m~s^{-1}]}$} & $2.0\text{e-03}$ & $1.32\text{e-03}$ & $1.32\text{e-04}$ & $1.32\text{e-05}$ & $1.32\text{e-06}$ & $1.32\text{e-07}$ \\ \hline
$\tau(\tau^{im})$ [Pa s]& 40 &  1.0e02 & 4.0e03 & 9.0e04& 4.0e05 & 1.0e06\\\hline
$\tau^{dr}$ [Pa s] & 8			 &	20		 & 800		& 1.8e04  & $8.0\text{e04}$ & $2.0\text{e05}$ \\ \hline 
$\tau_s$ 					 & $--$ 	 &	163.4  & 328.5 	& 4991  & $6.483\text{e04}$ & $6.083\text{e05}$\\ \hline 
    $\Delta t$ [s] & 7.3e-04 & 1.0e-03 & 5.6e-03 & 3.2e-02 & 1.8e-01& 1 \\ \hline
    $T_{end} \text{~[s]}$ ($\frac{Volume}{q \pi (d/2)^2}$) & $49.7$ &  $70.0$ & $409.7$ & $2.727\text{e03}$ & $1.2320\text{e04}$ & $1.3037\text{e05}$ \\
    $T_{end} \text{~[s]}$ (Model 1) & $54.0$ &  $216.0$ & $425.0$ & $2.380\text{e03}$ & $9.000\text{e03}$ & $3.120\text{e04}$ \\
    $T_{end} \text{~[s]}$ (Model 2) & $54.0$ &  $216.0$ & $460.0$ & $2.660\text{e03}$ & $1.000\text{e04}$ & $3.120\text{e04}$ \\
    $T_{end} \text{~[s]}$ (Model 3)&  $54.0$ &  $216.0$ & $460.0$ & $2.660\text{e03}$ & $1.000\text{e04}$ & $3.120\text{e04}$ \\\hline
  \end{tabular}
\end{table}

\begin{figure}[!htbp]
\begin{center}
  \subfigure[\label{fig:case_bc_1_6_1}Model 1 ]
  {\includegraphics[width=2.5in]{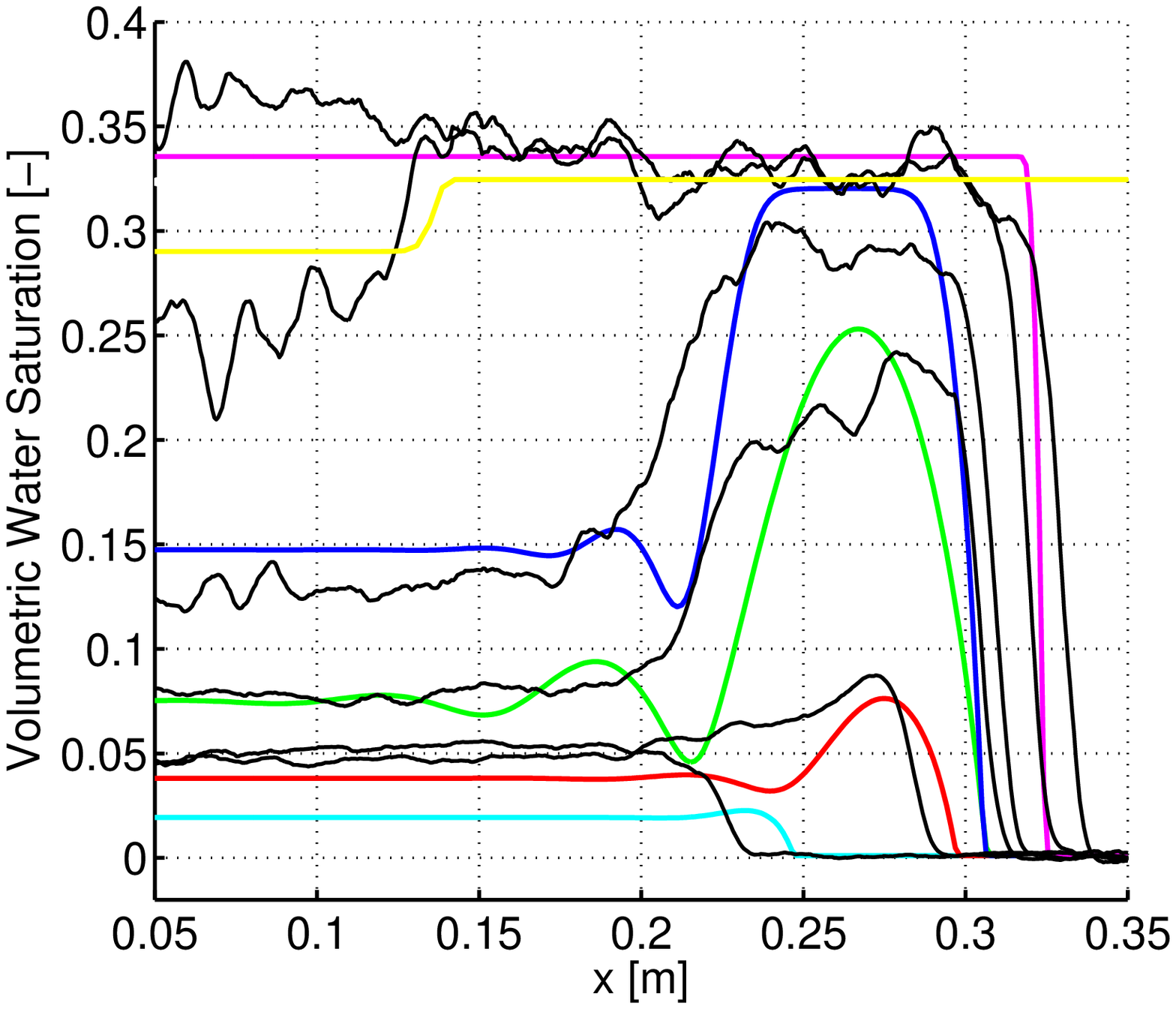}}\quad \quad
  \subfigure[\label{fig:case_bc_1_6_2}Model 2]
  {\includegraphics[width=2.5in]{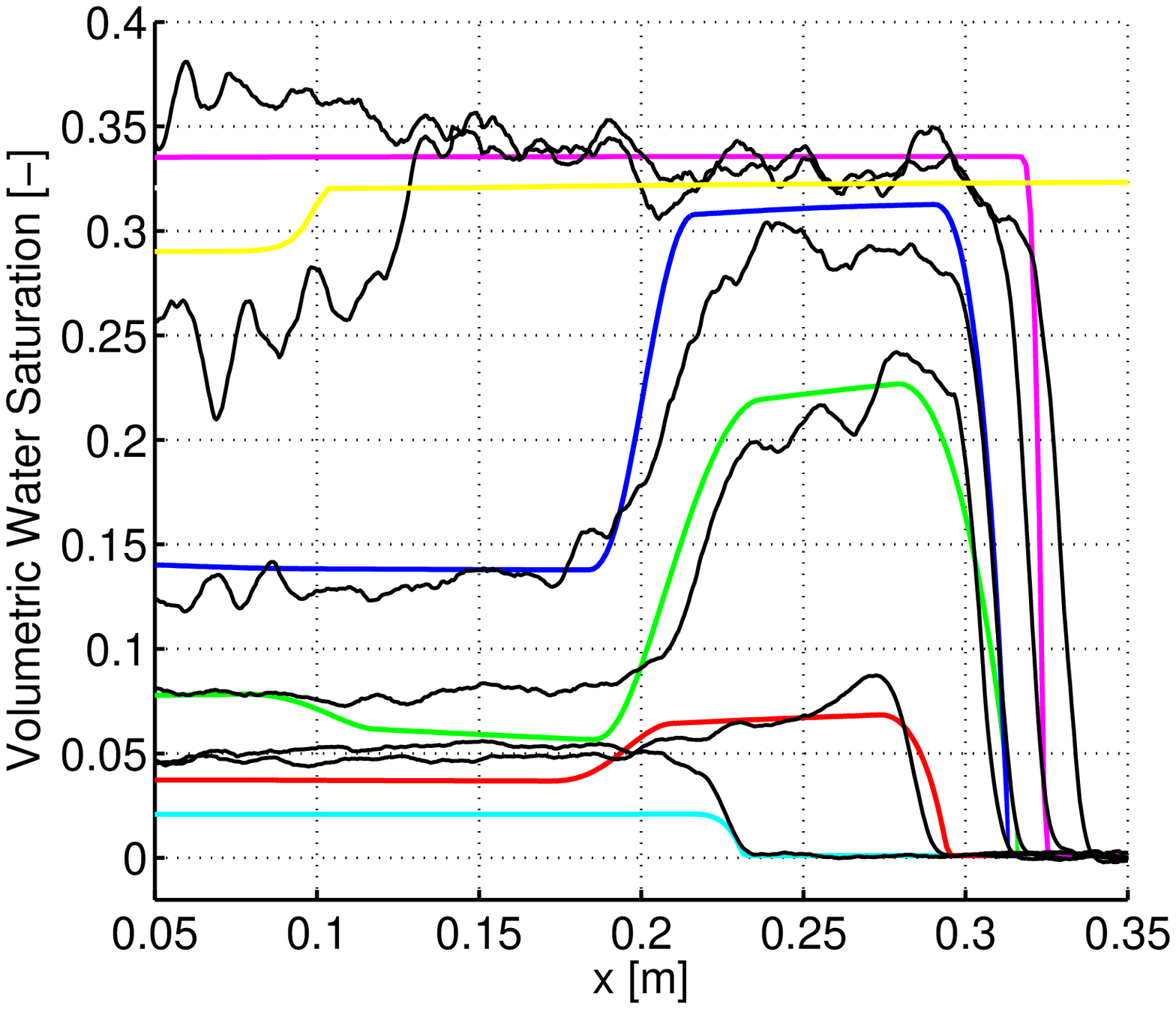}}\quad \quad
  \subfigure[\label{fig:case_bc_1_6_3}Model 3]
  {\includegraphics[width=2.5in]{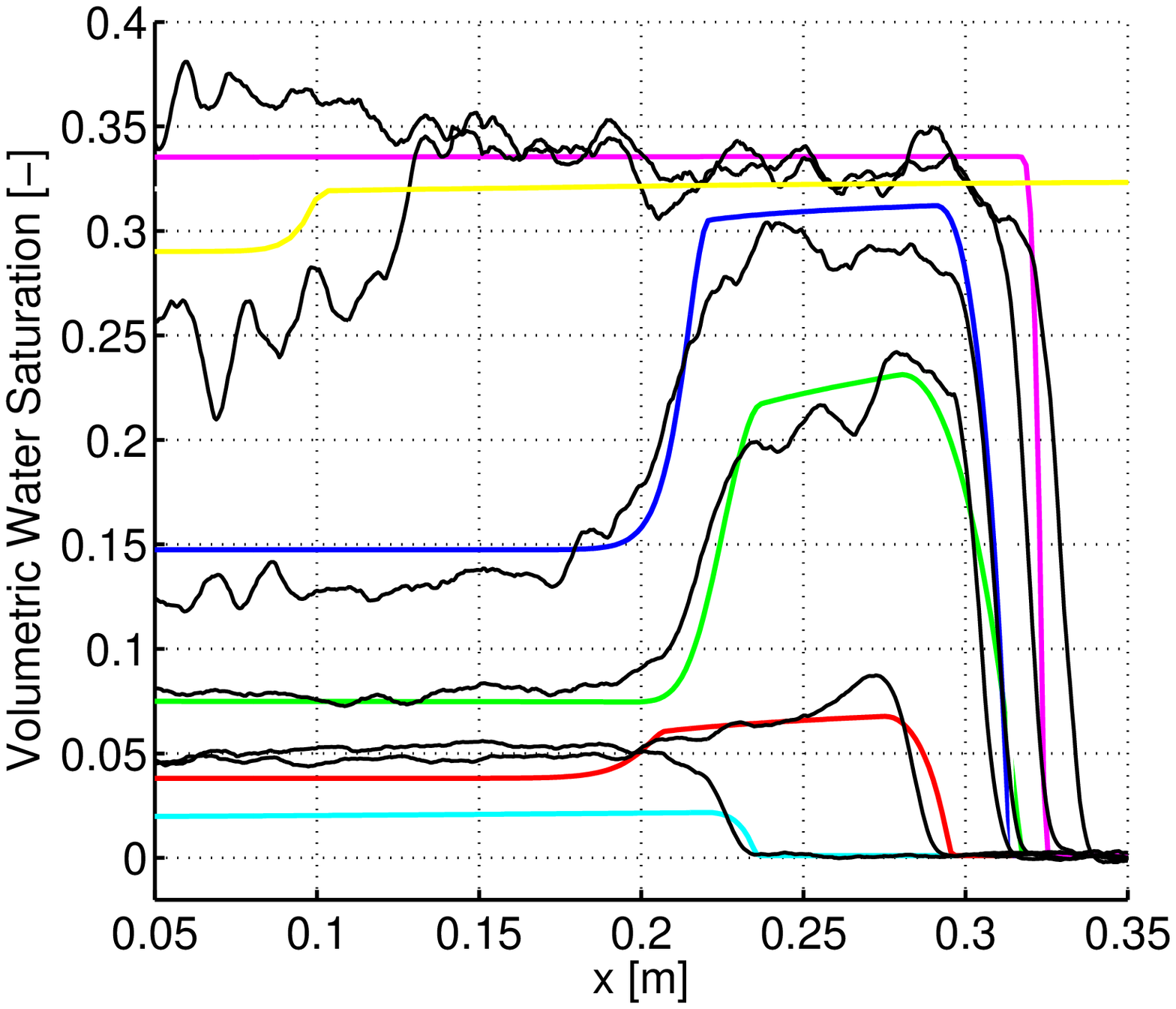}}\quad \quad
  \subfigure[ \label{fig:case_bc_2345_p_u}Phases pressure difference of Model 3 vs. water saturation]
  {\includegraphics[width=2.4in]{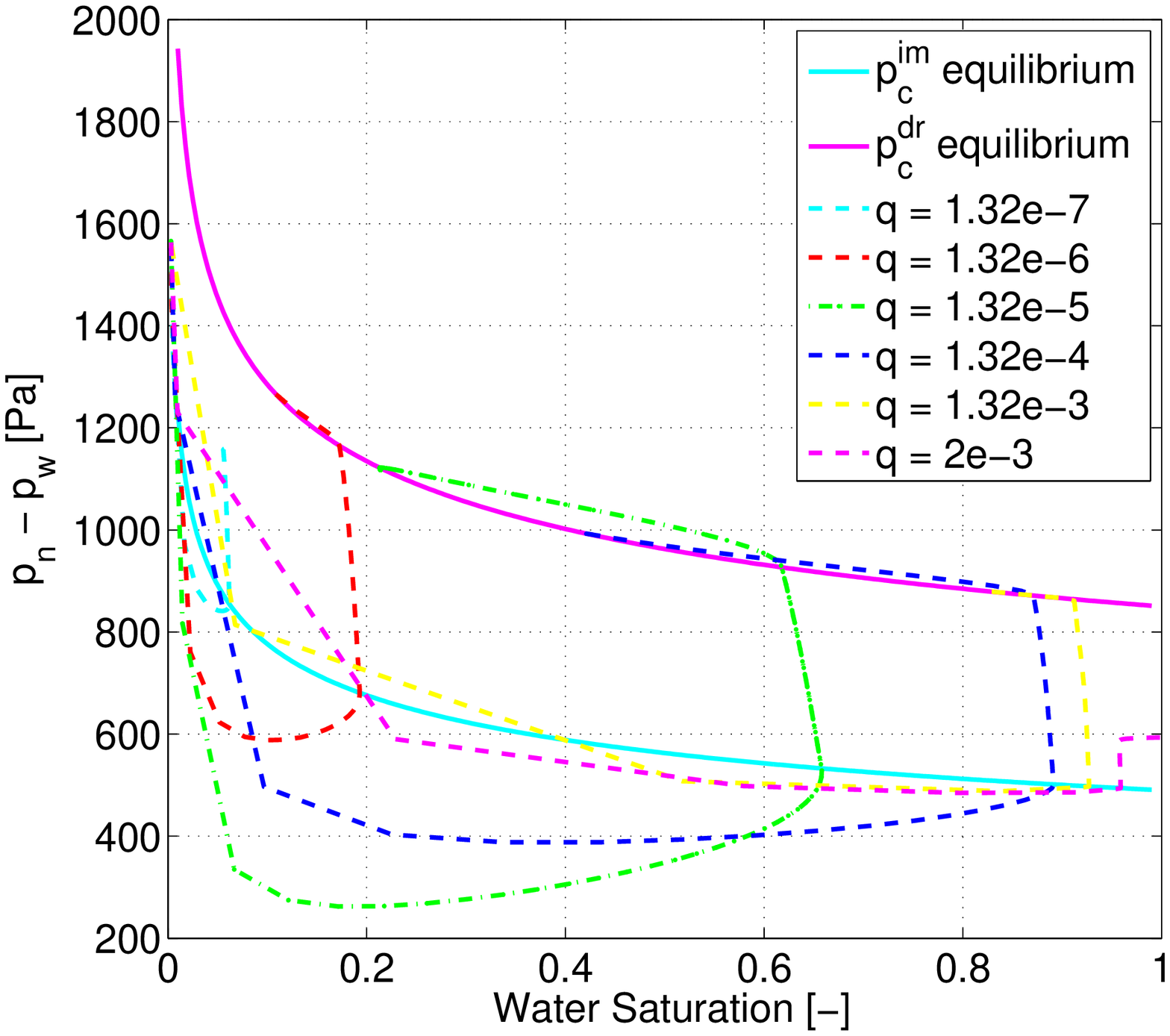}}
  \subfigure[\label{fig:case_bc_5}Solutions of Models 1, 2 and 3 when $q = 1.32\text{e-03}$]
  {\includegraphics[width=2.5in]{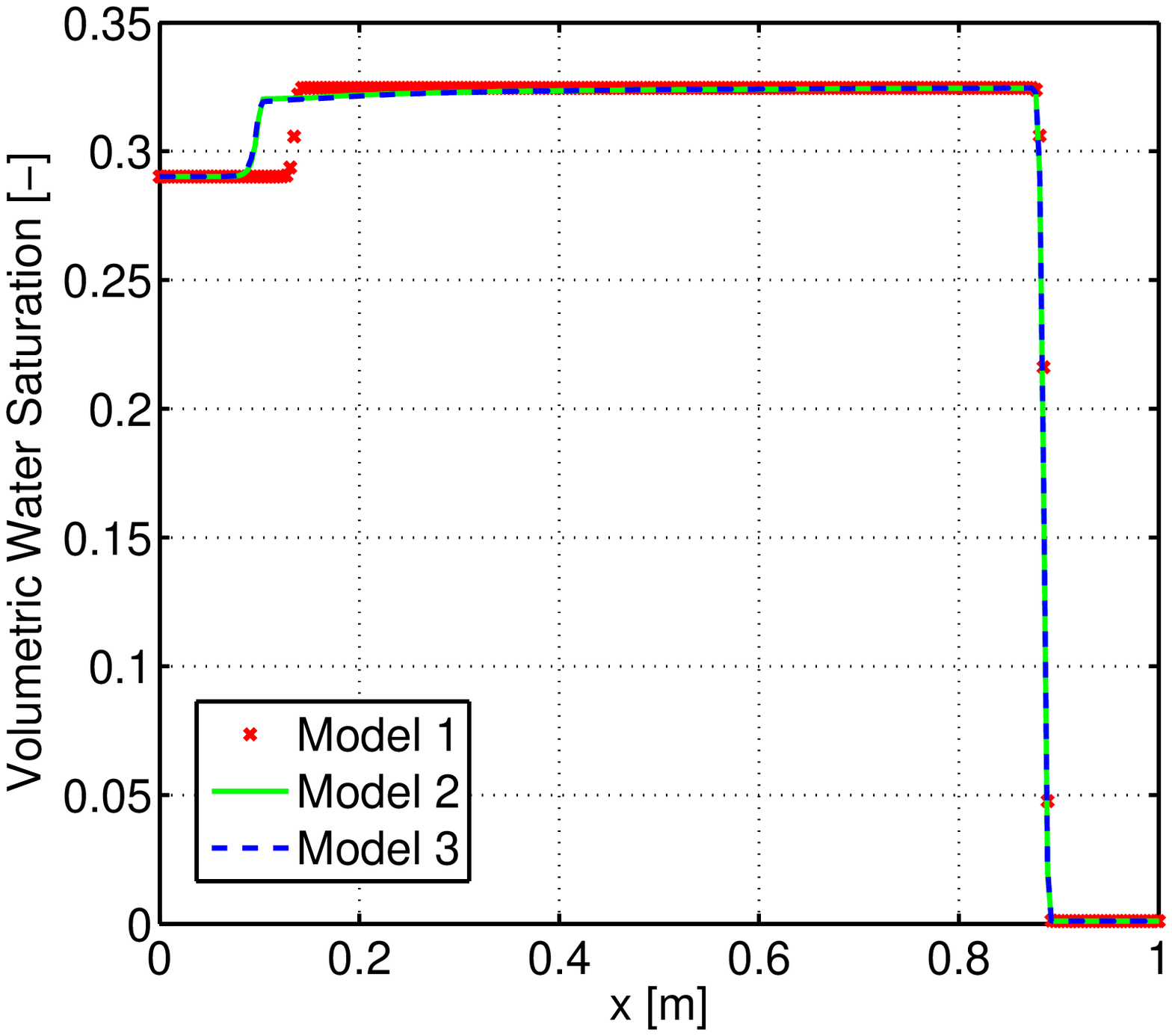}}\quad \quad
  \subfigure[\label{fig:case_bc_1}Solutions of Model 3 with $q = 1.32\text{e-07}$]
 {\includegraphics[width=2.5in] {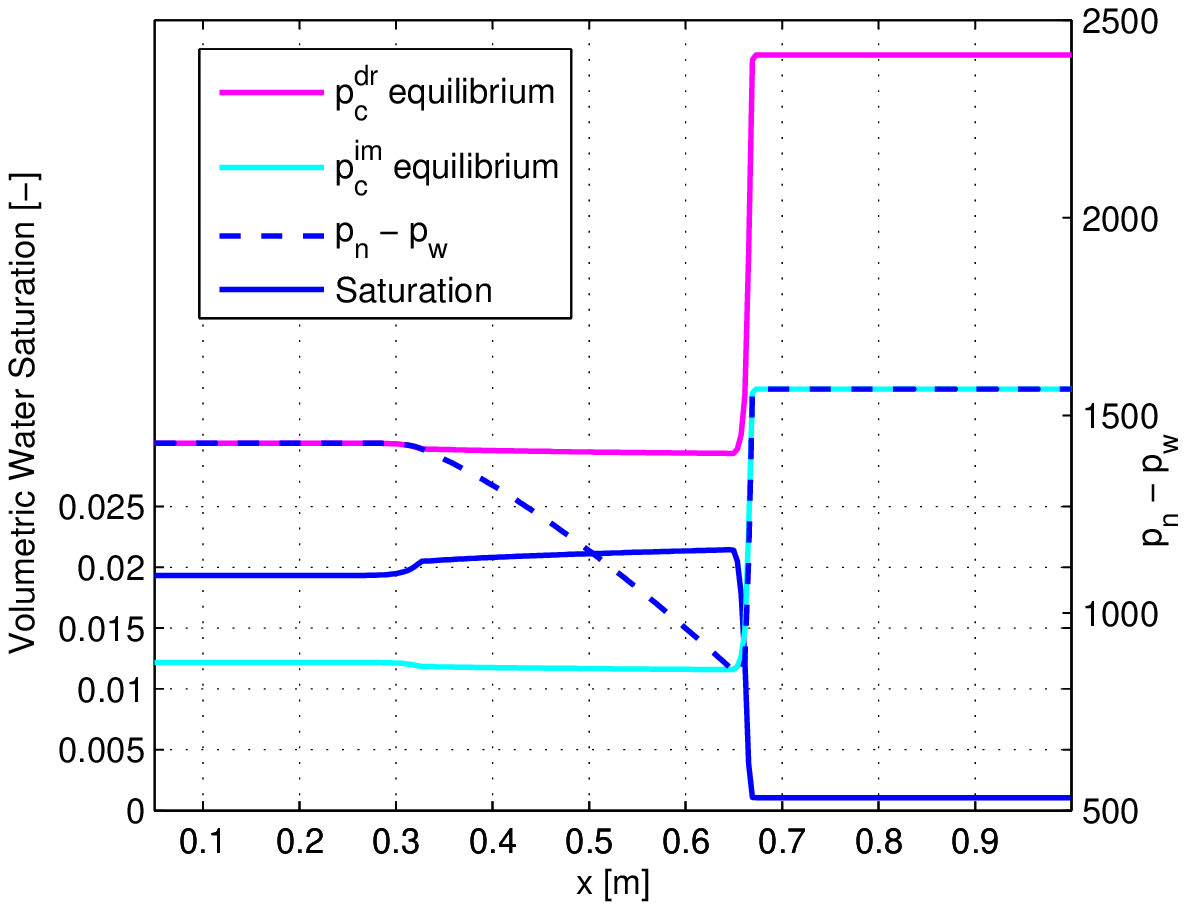}}
  \caption{Solutions obtained by Model 1, 2, 3 for different fluxes $q$. \newline
  (a), (b), (c) Cyan, red, green, blue, yellow and magenta lines are saturation profiles obtained when $q = 1.32\text{e-07}$, $1.32\text{e-06}, 1.32\text{e-05}, 1.32\text{e-04}, 1.32\text{e-03}, 2.0\text{e-03}$, respectively. Black lines denote experimental saturation. \newline
  (d) Cyan and magenta lines are the equilibrium imbibition and drainage capillary pressure curves obtained by the Brooks-Corey model. Cyan, red, green, blue, yellow and magenta dashed lines are obtained by Model 3 for different fluxes. \newline
  (e) Red cross, green line and dashed blue line are saturations obtained by Model 1, 2 and 3 when $q = 1.32\text{e-03}$.\newline
  (f) Blue, dashed blue, cyan and magenta lines denote numerical saturation, numerical capillary pressure, equilibrium imbibition and drainage pressure respectively.\newline
} \label{fig:case_bc_1_6}
\end{center}
\end{figure}

\begin{figure}[!htbp]
\begin{center}
  \subfigure[\label{fig:para}$\tau$, $T_{end}$ and $\Delta t$ for different fluxes ]
  {\includegraphics[width=2.5in]{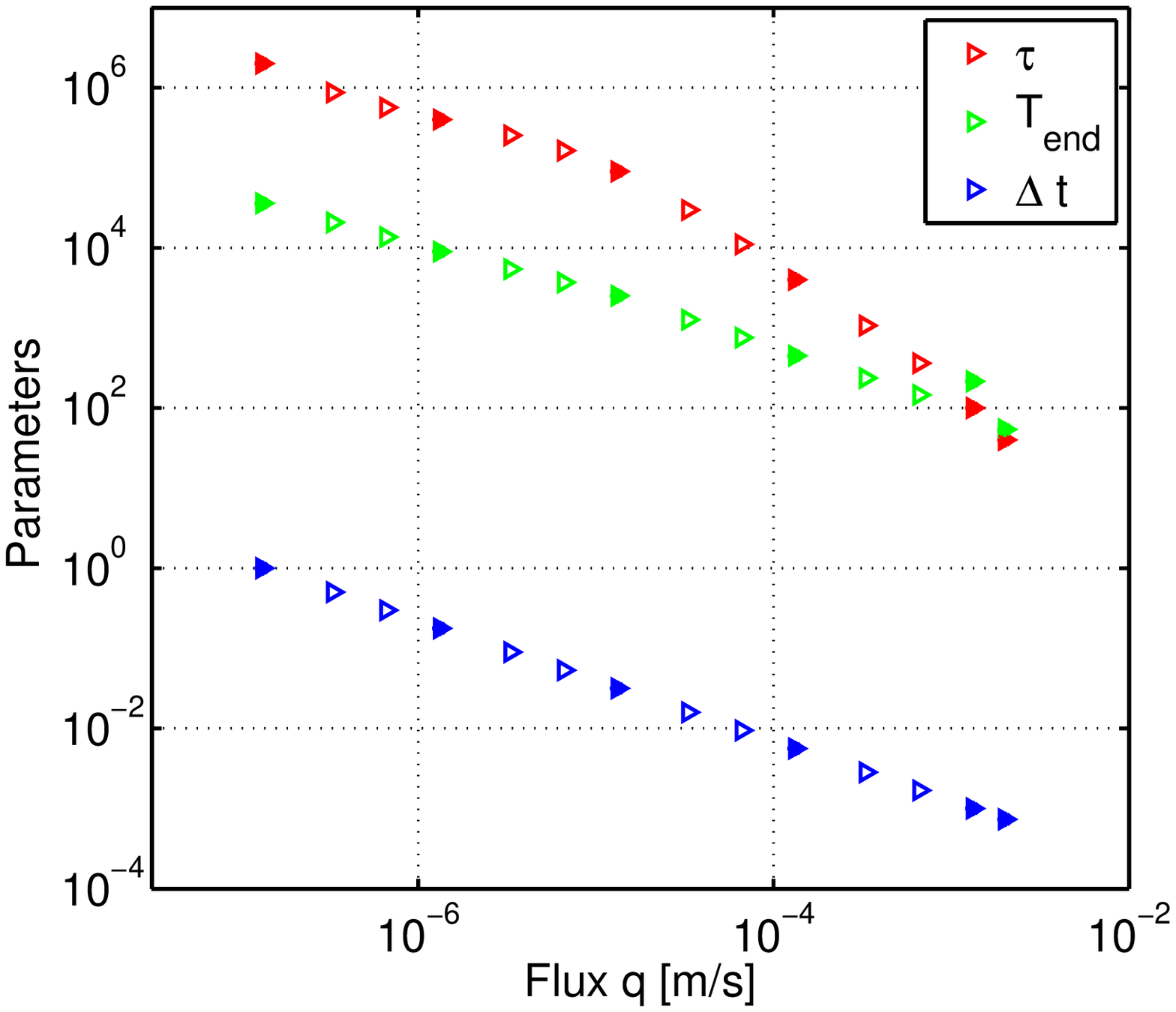}}\quad \quad
  \subfigure[\label{fig:tausaturation}$\tau$ as functions of tip and tail saturation]
  {\includegraphics[width=2.5in]{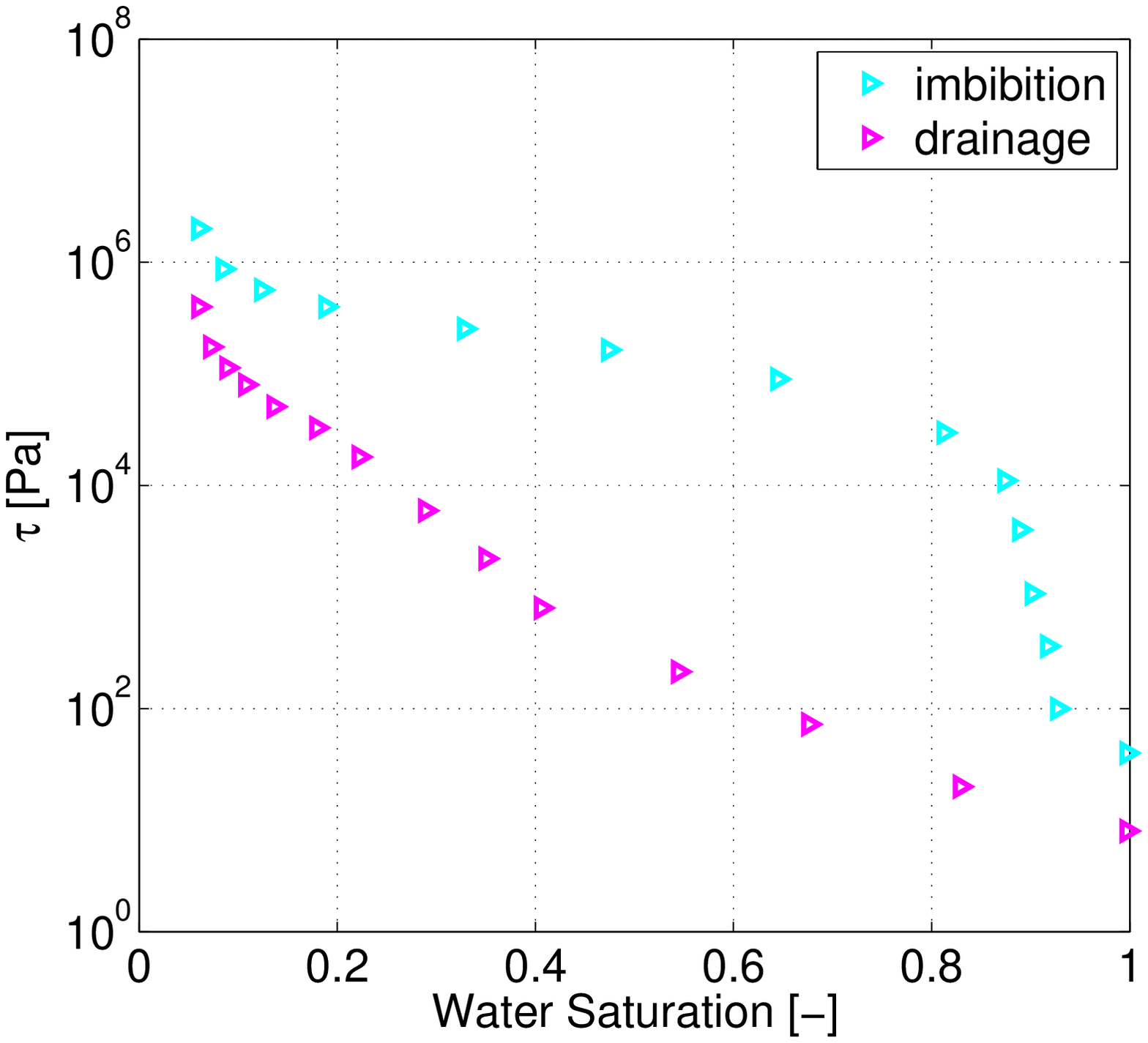}}
  \subfigure[\label{fig:tiptail}Tip and tail saturation vs. experiment]
  {\includegraphics[width=2.5in]{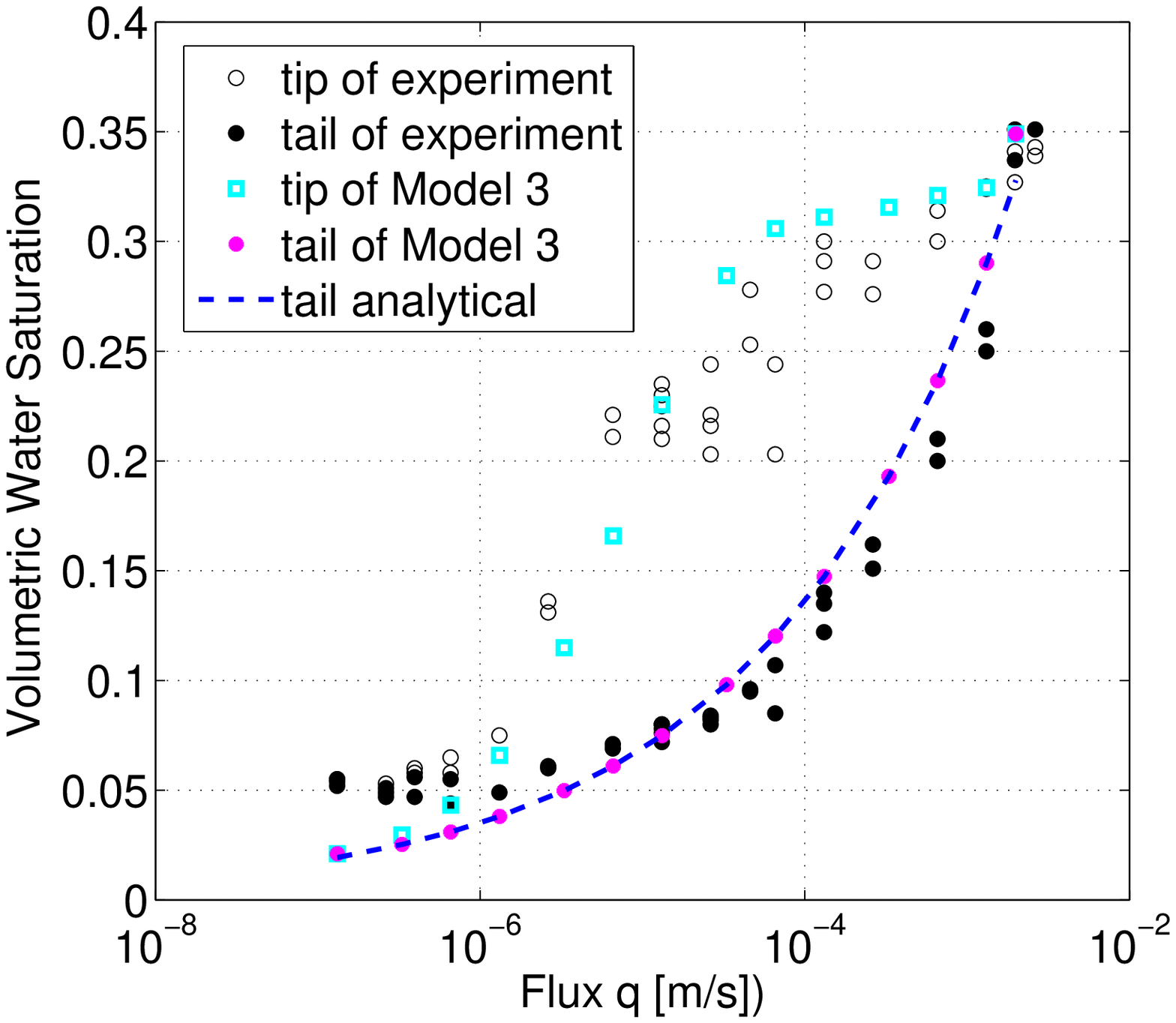}}\quad \quad
  % u1_6_tip_tail_linehystmnewana.eps is the old one
\subfigure[\label{fig:tiplength}Tip length as a function of flux]
 {\includegraphics[width=2.5in] {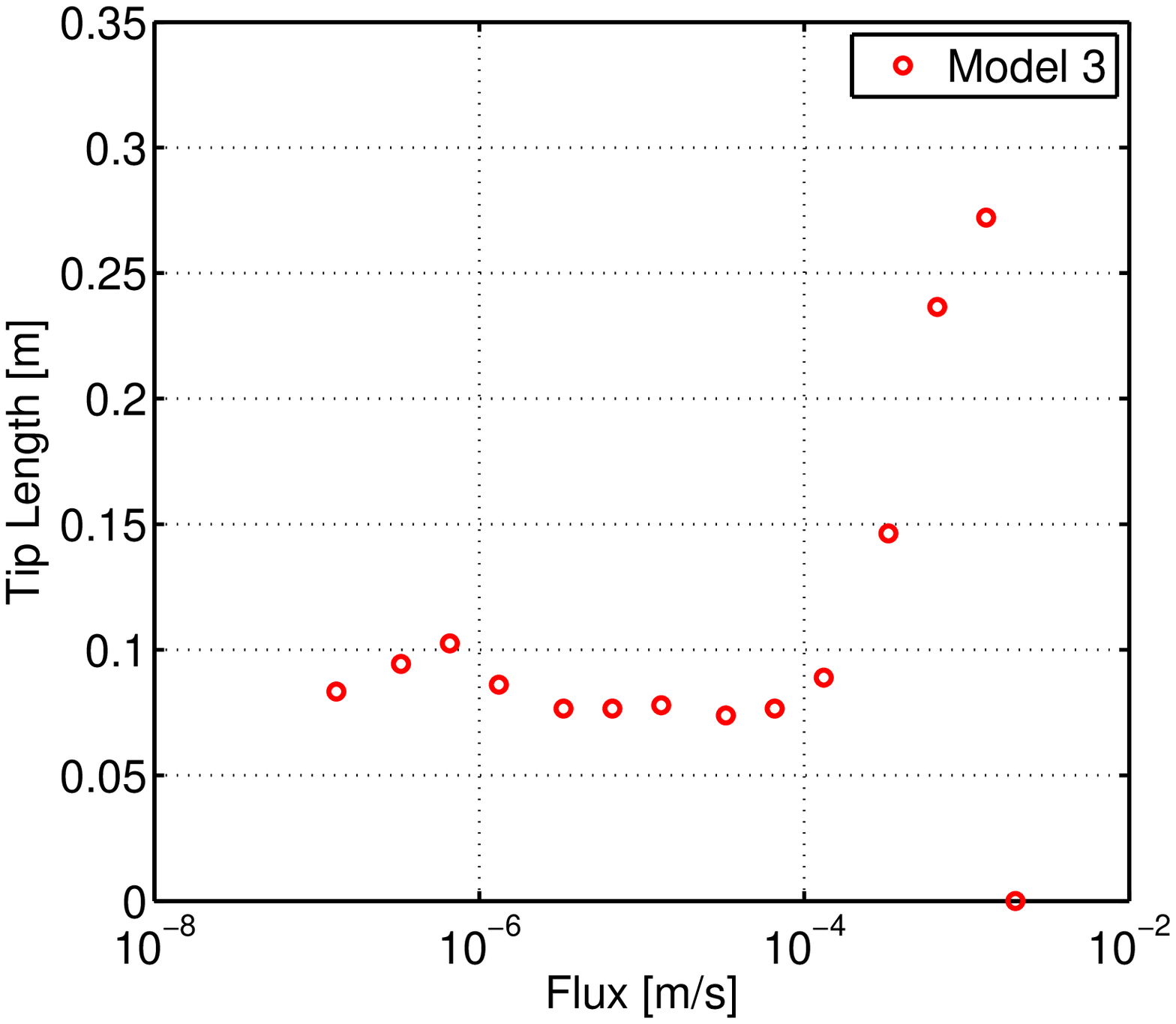}}
  \subfigure[\label{fig:pressatu}Phases pressure difference and saturation at the imbibition front and the tail]
  {\includegraphics[width=2.5in]{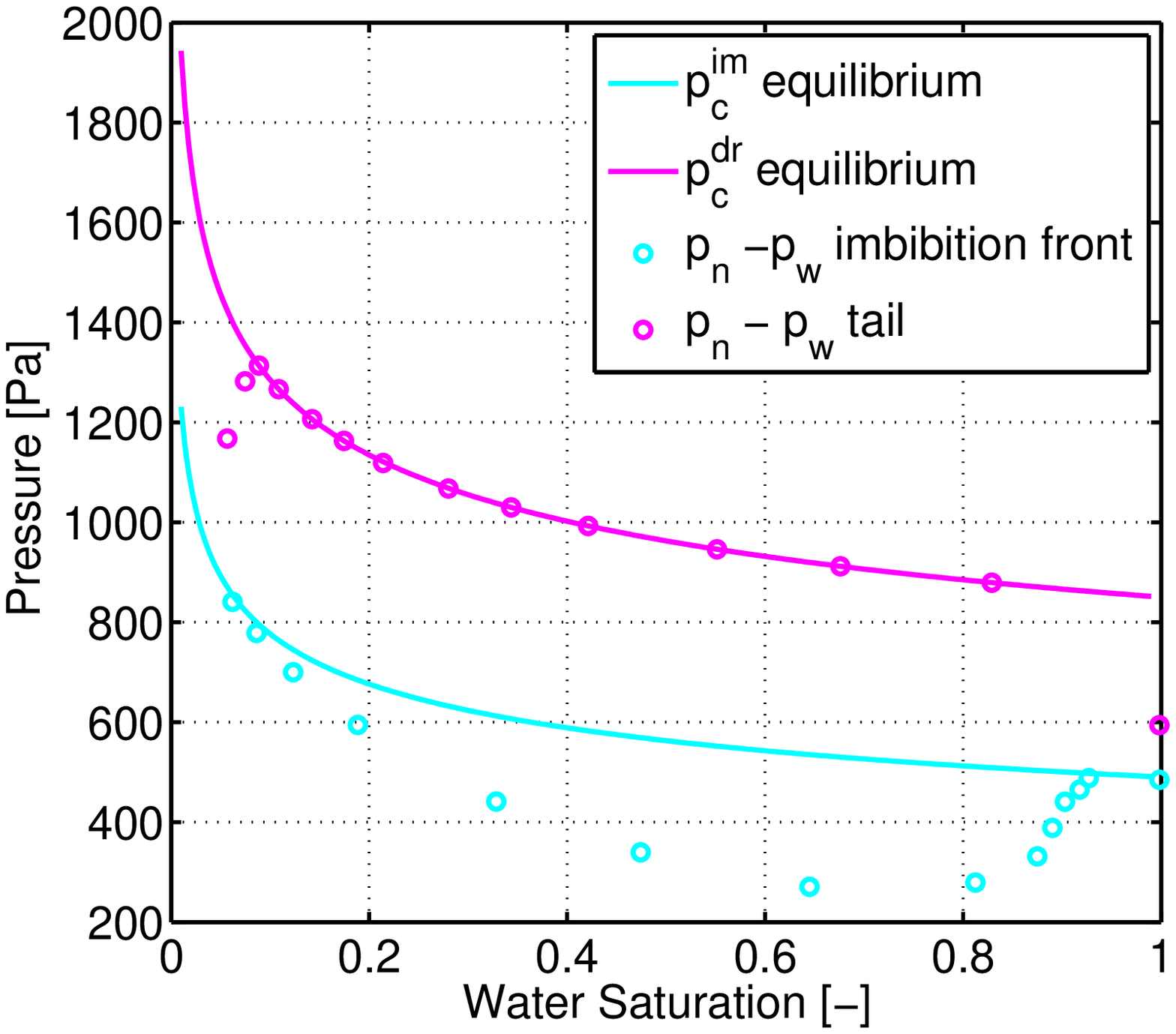}}\quad\quad
  \subfigure[\label{fig:presovershoot}Phases pressure difference and the overshoot for different fluxes]
  {\includegraphics[width=2.5in]{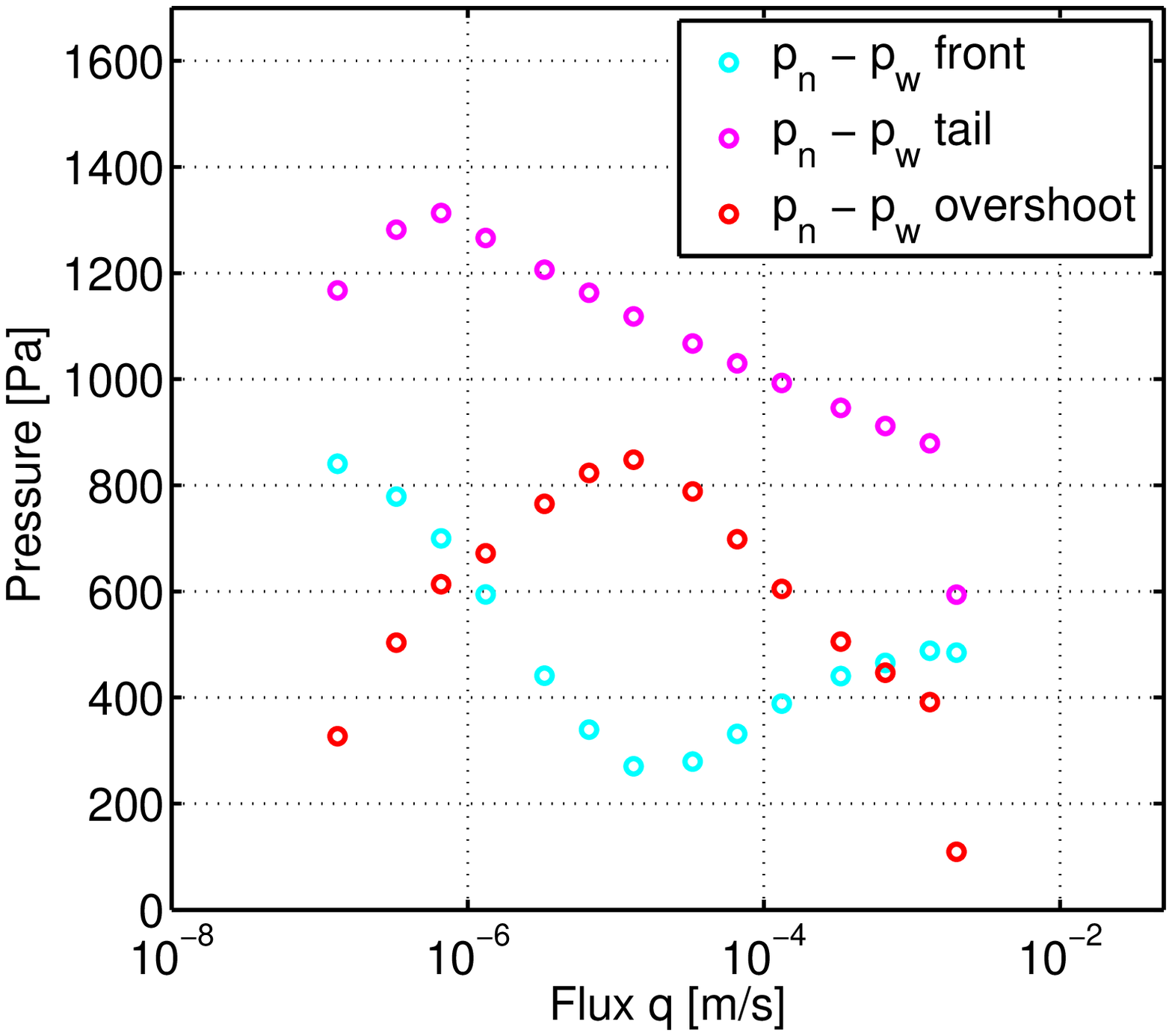}}
  \caption{Parameters and numerical results of Model 3.\newline
  (a) Red, green and blue triangles denote values of $\tau$, $T_{end}$ and $\Delta t$. Solid triangles denote the parameters used in Fig. \ref{fig:case_bc_1_6_3}, open triangles are obtained by interpolating solid triangles. These values are used in (b), (c), (d), (e), (f).\newline
   (b) Cyan and magenta lines denote the values of $\tau$ at the imbibition front and the tail.\newline
   (c) Experimental data is from \cite{dicarlo2004experimental}. Cyan squares and magenta solid circles are the tip and tail saturation obtained by Model 3. Blue line denotes the tail saturation obtained by Eq. (\ref{eqn:boundsatuex}).\newline
   (d) Red circles are tip lengths for different fluxes.\newline
   (e) Cyan and magenta lines are the equilibrium imbibition and drainage capillary pressures obtained by the Brooks-Corey model. Cyan and magenta circles are phases pressure differences at the imbibition front and the tail.\newline
   (f) Cyan and magenta circles denote phases pressure difference at the imbibition front and the tail, red circles are the overshoot.\newline
} \label{fig:case_bc_all}
\end{center}
\end{figure}

To our best knowledge, the $\tau$ values are not known for the 20/30 sand, thus we have to first try different values of $\tau$ and then find the best match with the experiments. \cite{dicarlo2004experimental} observed at the highest $2.0\text{e-03}$ and lowest $1.32\text{e-07}$ fluxes the saturation profiles are monotonic with distance and no saturation overshoot is observed. In order to find suitable values of $\tau$ for the highest and lowest fluxes, in Fig. \ref{fig:para} we plot $\tau$ as a function of $q$ using a log-log diagram. Realizing the near log-log relationship between $\tau$ and $q$, we set $\tau = 40$ for $q = 2.0\text{e-03}$, $\tau = 2.0\text{e06}$ for $q = 1.32\text{e-07}$. The number of nodes used in space is $N = 256$, the values of $\tau$($\tau^{im}$), $\tau^{dr}$, $\tau_s$, time steps as well as the end times are presented in Table \ref{tab:endtime}. As can be seen the end times for simulations are near to the calculated times except when $q = 1.32\text{e-07}$, we will explain this later. For all three models, when the flux $q$ is $1.32\text{e-03}$, the imbibition front moves quickly while the change in hysteretic capillary pressure is slow. In order to show the overshoot saturation phenomenon, we have to enlarge the interval to $[0, 1]$. In Fig. \ref{fig:case_bc_1_6} we compare the numerical solutions with the experiments.

Figs. \ref{fig:case_bc_1_6_1}, \ref{fig:case_bc_1_6_2} and \ref{fig:case_bc_1_6_3} show that all three models can obtain saturation overshoots for $q = 1.32\text{e-06}$, $q = 1.32\text{e-05}$, $q = 1.32\text{e-04}$, $q = 1.32\text{e-03}$. From Fig. \ref{fig:case_bc_1_6_1} we can see oscillations behind the drainage fronts when $q = 1.32\text{e-06}$, $q = 1.32\text{e-05}$ and $q = 1.32\text{e-04}$, while Fig. \ref{fig:case_bc_1_6_2} only present weaker oscillations for $q = 1.32\text{e-04}$ and $1.32\text{e-05}$ and there is no oscillation behind the drainage fronts in Fig. \ref{fig:case_bc_1_6_3}. Althouth Table \ref{tab:endtime} shows that for $q = 1.32\text{e-04}$, $1.32\text{e-05}$ and $¡ì.32\text{e-06}$, the $\tau^{dr}$ values are slightly larger than $\tau_s$, no oscillation appears behind the drainage fronts. This may be cuased by the hystersis in the capillary pressure, because in Fig. \ref{fig:case_bc_1_6_2} the oscillations are weaker than Fig. (\ref{fig:case_bc_1_6_1}) even without hysteresis in dynamic capillary coefficient. 

In Fig. \ref{fig:case_bc_2345_p_u} we plot the relationship between phases pressure difference and saturation obtained by Model 3 for all fluxes. At the imbibition front, the phases pressure difference is smaller than the equilibrium imbibition pressure. Behind the front, the pressure-saturation follows the equilibrium drainage pressure while finally stops at the tail saturation. This figure shows similar pressure-saturation behaviour as the experiments presented in Fig. 6 in \cite{dicarlo2007capillary}.  As can be seen, at the lowest flux, significant pressure overshoot can still be obtained by Model 3. This phenomenon was also observed in the experiment in \cite{dicarlo2007capillary}. \cite{selker1992fingered} shows that saturation overshoot is associated with pressure overshoot. Thus we guess even that at low flux, Model 3 can still produce saturation overshoot. Let the space interval be $[0, 1]$ and $T_{end} = 96000$, the saturation and pressure profiles computed by Model 3 are presented in Fig. \ref{fig:case_bc_1}. It shows at $q = 1.32\text{e-07}$ very small saturation overshoot appears.

Fig. \ref{fig:case_bc_1_6_1}, \ref{fig:case_bc_1_6_2}, \ref{fig:case_bc_1_6_3} also show that the computed saturations at the left boundary differ from the experiments especially when $q = 1.32\text{e-07}$. We ascribe this to the limitation of the Brooks-Corey model. Assuming that after a long time, the saturation at the boundary reaches the equilibrium state, we set $\frac{\partial S_w}{\partial t}|_{x = 0} = 0, \frac{\partial S_w}{\partial x}|_{x = 0} = 0$. Then we obtain
\begin{align}\label{eqn:boundsatu}
  q f(S_w) + \lambda_n(S_w) f(S_w) (\rho_w - \rho_n)g = q.
\end{align}
Using the parameters for imbibition process in Table \ref{tab:2030sand} and the Brooks-Corey model in Table \ref{tab:model}, from Eq. (\ref{eqn:boundsatu}) we get
\begin{align}
  \label{eqn:boundsatuex}
  S_w = \left( \frac{\mu_w q }{K (\rho_w - \rho_n)g}\right)^{\frac{\lambda}{2 + 3\lambda}}.
\end{align}
The saturations obtained by Eq. (\ref{eqn:boundsatuex}), the numerical simulations and experiments at $x = 0$ are presented in Table \ref{tab:boundsatu}. The measured volumetric water saturation at $q = 1.32\text{e-07}$ is twice as high as the analytical one. Thus, the end time of the numerical simulation is much shorter than the calculated time.
\begin{table}
  \caption{\label{tab:boundsatu}Saturations at $x = 0$ obtained by Eq. (\ref{eqn:boundsatuex}) and numerical simulations.}
  \center
  \begin{tabular}{|c|ccccc|}\hline
  {}&\multicolumn{5}{c|}{Brooks-Corey Model}\\\cline{2-6}
\raisebox{1.6ex}[0pt]{$q\mathrm{~[m s^{-1}]}$} & $\theta$ (Experiment) & $\theta$ (Analytical) &$\theta$ (Model 1)& $\theta$ (Model 2) & $\theta$ (Model 3)\\\hline
    2.0e-03  & 0.3500 & 0.3279 & 0.3289 & 0.3491 & 0.3494\\
    1.32e-03 & 0.2665 & 0.2902 & 0.2902 & 0.2902 & 0.2902\\
    1.32e-04 & 0.1250 & 0.1474 & 0.1474 & 0.1421 & 0.1474\\
    1.32e-05 & 0.0790 & 0.0749 & 0.0749& 0.0777 & 0.0749\\
    1.32e-06 & 0.0500 & 0.0380 & 0.0380 & 0.0377&0.0380\\
    1.32e-07 & 0.0450 & 0.0193 & 0.0193 & 0.0211 & 0.0198 \\ \hline
  \end{tabular}
\end{table}

Since Fig. \ref{fig:case_bc_1_6_3} shows more realistic profiles than Fig. \ref{fig:case_bc_1_6_1} and \ref{fig:case_bc_1_6_2}, we will focus on this model and apply more fluxes to test its effectiveness. In Fig. \ref{fig:para} we plot $\tau$, $T_{end}$, $\Delta t$ and $q$ used in Fig. \ref{fig:case_bc_1_6} with solid triangles. Then we use logarithmic interpolation to get the values of $\tau$, $T_{end}$ and $\Delta t$ for intermediate fluxes, these parameters are plotted using open triangles. Fig. \ref{fig:tausaturation} plots the values of $\tau$ as functions of tip and tail saturations. For both imbibition and drainage, the values of $\tau$ increase as water saturations decrease. This trend seems agree with the measured data in \cite{manthey2005macro,das2012dynamic,mirzaei2013experimental}.

In Fig. \ref{fig:tiptail} the computed tip and tail saturations from Model 3 are compared with the measured data in \cite{dicarlo2004experimental}. As the flux increases, the tip saturation increases very fast for flux value in interval $[1.0\text{e-06}, 1.0\text{e-04}]$, while the tip saturation increases slowly when flux $q$ is above $1.32\text{e-04}$. For tail saturations, both the experimental data and computed results follow the analytical curve given by Eq. (\ref{eqn:boundsatuex}) when flux is bigger than $1.0\text{e-06}$.

Fig. \ref{fig:tiplength} plots the tip length versus flux obtained by Model 3. For flux values between $1.0\text{e-05}$ and $1.0\text{e-03}$, the tip length increases monotonically with the flux. This trend matches Fig. 12 in \cite{dicarlo2004experimental}.

Fig. \ref{fig:pressatu} presents the phases pressure differences at the imbibition front and the tail obtained by Model 3. For intermediate fluxes, the phases pressure differences at the tail follow the equilibrium drainage capillary pressure, while the phases pressure differences at the imbibition front are below the equilibrium imbibition capillary pressure as a result of the dynamic capillary pressure effect.

\cite{dicarlo2007capillary} defined the overshoot in capillary pressure and the overshoot of phases pressure difference is given as
\begin{align}
  \label{eqn:overshootcapi}
  \mathrm{Overshoot}(p_n - p_w) = \mathrm{Tail}(p_n - p_w) - \mathrm{Front}(p_n - p_w).
\end{align}
Fig. \ref{fig:presovershoot} plots the phases pressure differences as well as the overshoots for different fluxes. The phases pressure differences at the imbibition front are higher at low flux values than at high flux values. At $q \approx  1\text{e-05}$, the phases pressure difference at the imbibition front reaches a minimum while the pressure overshoot reaches a maximum.
%  {\includegraphics[width=2.5in]{pic/u1_6_tip_tail_line.eps}}\quad
\section{Conclusion}
In this study we applied the Castillo-Grone's mimetic operators and the implicit trapezoidal rule to solve two-phase flow models including dynamic capillary pressure (with constant and hysteretic coefficient) and capillary pressure hysteresis in porous media. Numerical simulations show that the second-order mimetic operators mimics the Green-Gauss-Stokes theorem with high accuracy for all three models. The hysteretic dynamic capillary pressure model with capillary pressure hysteresis produce realistic saturation overshoot and pressure overshoot phenomena as observed in \cite{dicarlo2004experimental,dicarlo2007capillary}.

\begin{acknowledgements}
The research of H. Zhang was funded by the China scholarship Council (No. 201503170430). We thank Prof. S. Majid Hassanizadeh for useful discussions. Moreover, the valuable suggestions and comments from the anonymous reviewers are highly appreciated.
\end{acknowledgements}

% BibTeX users please use one of
\bibliographystyle{aps-nameyear}      % American Physical Society (APS) style, author-year citations
\bibliography{reference}

\end{document}